\documentclass[a4paper]{article}

\usepackage{authblk}
\usepackage{graphicx}
\usepackage{amssymb}
\usepackage{appendix}

\newcommand{\tr}{\mathrm{tr}}
\newcommand{\diag}{\mathrm{diag}}

\newenvironment{mat}[1]{\left[\begin{array}{#1}}{\end{array}\right]}
\newcommand{\bmx}[1]{\begin{mat}{#1}}
\newcommand{\emx}{\end{mat}}

\newcommand{\gssa}[4]{\mbox{\boldmath $#1$}_{#2}^{#3}(#4)}
\newcommand{\ga}[2]{\mbox{\boldmath $#1$}(#2)}
\newcommand{\gss}[3]{\mbox{\boldmath $#1$}_{#2}^{#3}}
\newcommand{\g}[1]{\mbox{\boldmath $#1$}}

\newcommand{\sbm}[1]{\mbox{\scriptsize $\g{#1}$}}

\newcommand{\pref}[1]{(\ref{#1})}

\begin{document}

\title{Preamble-based Channel Estimation in OFDM/OQAM Systems: A Time-Domain Approach\thanks{This work was supported by an FP7 ICT grant, project EMPhAtiC (www.ict-emphatic.eu).}}

\author{\emph{Eleftherios Kofidis}}
\affil{Department of Statistics and Insurance Science, University of Piraeus, 80, Karaoli \& Dimitriou str., 185~34 Piraeus, Greece (E-mail: kofidis@unipi.gr) \\and\\
Computer Technology Institute (CTI), 26~500 Patras, Greece}

\date{}

\maketitle

\begin{abstract}
Filter bank-based multicarrier (FBMC) systems based on offset quadrature amplitude modulation (FBMC/OQAM) have recently attracted increased interest (in applications including DVB-T, cognitive radio, and powerline communications) due to their enhanced flexibility, higher spectral efficiency, and better spectral containment compared to conventional OFDM. 
FBMC/OQAM suffers, however, from an imaginary inter-carrier/inter-symbol interference that complicates signal processing tasks such as channel estimation. Most of the methods reported thus far in the literature rely on the assumption of (almost) flat subchannels to more easily tackle this problem, with the aim of addressing it in a way similar to OFDM. 
However, this assumption may be often quite inaccurate, due to the high frequency selectivity of the channel and/or the small number of subcarriers employed to cope with frequency dispersion in fast fading environments. In such cases, severe error floors are exhibited at medium to high signal-to-noise ratio (SNR) values, that cancel the advantage of this modulation over OFDM. Moreover, the existing methods provide estimates of the subchannel responses, most commonly in the frequency domain. The goal of this paper is to revisit this problem through an alternative formulation that focuses on the estimation of the channel impulse response itself and makes no assumption on the degree of frequency selectivity of the subchannels. The possible gains in estimation performance offered by such an approach are investigated through the design of optimal (in the mean squared error sense) preambles, of both the full and sparse types, and of the smallest possible duration of only one pilot FBMC symbol. Existing preamble designs for flat subchannels are then shown to result as special cases. The case of longer preambles, consisting of two consecutive pilot FBMC symbols, is also analyzed. Simulation results are presented, for both mildly and highly frequency selective channels, that demonstrate the significant improvements in performance offered by the proposed approach over both OFDM and the optimal flat subchannel-based FBMC/OQAM method. Most notably, no error floors appear anymore over a quite wide range of SNR values. 
\end{abstract}

\section{Introduction}
\label{sec:intro}

Filter bank-based multicarrier (FBMC) systems based on offset quadrature amplitude modulation (known as FBMC/OQAM or OFDM/OQAM) have recently attracted increased interest (in applications including DVB-T~\cite{ar11}, cognitive radio~\cite{akpf07,PHYDYAS}, and powerline communications~\cite{als11}) due to their enhanced flexibility, higher spectral efficiency, and better spectral containment compared to conventional OFDM. Notably, there is no need to insert a guard interval such as the cyclic prefix (CP) common in OFDM. 
FBMC/OQAM suffers, however, from an imaginary inter-carrier/inter-symbol interference that complicates signal processing tasks such as channel estimation~\cite{jlr03}. Most of the methods reported thus far in the literature rely on the assumption of (almost) flat subchannels to more easily tackle this problem, with the aim of addressing it in a way similar to OFDM; see~\cite{kkrt13} and references therein. 
However, this assumption may be often quite inaccurate, due to the high frequency selectivity of the channel~\cite{sir08,bnn12} and/or the small number of subcarriers employed to cope with frequency dispersion in fast fading environments~\cite{gmm08}. In such cases, and due to the residual interference generated by the inaccuracies in the system model adopted, severe error floors are exhibited at medium to high signal-to-noise ratio (SNR) values, which cancel the advantage of this modulation over CP-OFDM. Moreover, the existing methods provide estimates of the subchannel responses (e.g., \cite{bnn12}), most commonly in the frequency domain~\cite{kkrt13}. The goal of this paper is to revisit this problem through an alternative formulation that focuses on the estimation of the channel impulse response itself and makes no assumption on the degree of frequency selectivity of the subchannels.\footnote{A similar approach (however not addressing the training design problem) was recently proposed in~\cite{kqj14}.} The possible gains in estimation performance offered by such an approach are investigated through the design of optimal (in the mean squared error (MSE) sense) preambles, of both the full and sparse types~\cite{kkrt09}, consisting of only one pilot FBMC symbol. Existing preamble designs for flat subchannels~\cite{kkrt09,kkrt10,kkrt13} are then shown to result as special cases. The case of longer preambles, consisting of two consecutive pilot FBMC symbols, is also analyzed. Simulation results are presented, for both mildly and highly frequency selective channels, that demonstrate the significant improvements in performance and robustness to channel dispersion offered by the proposed approach over both CP-OFDM and the optimal flat subchannel-based FBMC/OQAM method. Most notably, no error floors appear anymore over a quite wide range of SNR values. 

The rest of the paper is organized as follows. An alternative formulation of the FBMC/OQAM system in the presence of channel and noise is developed in Section~\ref{sec:formulation}. The aim is to develop an input-output expression explicitly in terms of the channel impulse response (CIR), and most importantly \emph{independently} of the time dispersion of the channel relative to the filter bank size. Section~\ref{sec:CE} then goes ahead with deriving the CIR estimator and paving the way to preamble design. The latter is addressed in Section~\ref{sec:optimal}, for both \emph{full} (i.e., all subcarriers carrying pilots) and \emph{sparse} (i.e., isolated pilot tones surrounded by nulls) preambles. The criterion is the minimization of the channel estimation MSE subject to a given energy budget at the output of the synthesis filter bank (modulator). Simulation results are reported that demonstrate the effectiveness of the proposed approach in coping with the inteference effect even in highly frequency selective channels and over a wide range of SNRs. Preambles concisting of more than one pilot FBMC symbol are considered in Section~\ref{sec:longer}. 
The case of flat subchannels is recalled in the appendix, where the known solutions for full and sparse preambles in this scenario~\cite{kkrt09} are re-derived as special cases of the results obtained previously. For flat subchannels, frequency smoothing recently developed in~\cite{krbh13} is extended to incorporate knowledge of the inter-subcarrier noise correlation and the performance improvement it can offer with full preambles is exemplified. 

\medskip

\noindent
\emph{Notation.} Lower and upper case bold letters will be used to denote vectors and matrices, respectively. Superscripts $^{T}$ and $^{H}$ will respectively signify transposition and conjugate (Hermitian) transposition. Complex conjugation will be denoted by $^{*}$. $\gss{0}{r\times s}{}$ will be the all zeros matrix of dimensions $r\times s$. $\gss{I}{s}{}$ is the $s$th-order identity matrix. The $r\times 1$ vector of all ones will be denoted by $\gss{1}{r}{}$. The diagonal matrix with the entries of the vector $\g{x}$ on its main diagonal will be denoted by $\diag(\g{x})$. 

\section{Time-domain Formulation}
\label{sec:formulation}

With the usual notation adopted, the (baseband) output of an FBMC/OQAM synthesis filter bank (SFB) can be written as
\begin{equation}
s(l)=\sum_{m=0}^{M-1}\sum_{n}d_{m,n}g_{m,n}(l),
\label{eq:SFB}
\end{equation}
where $(m,n)$ refers to the $m$th subcarrier and the $n$th FBMC symbol, $d_{m,n}$ are \emph{real} OQAM symbols, $M$ is the (even) number of subcarriers, and 
\[
g_{m,n}(l)=g\left(l-n\frac{M}{2}\right)e^{j\frac{2\pi}{M}m\left(l-\frac{L_{g}-1}{2}\right)}e^{j\varphi_{m,n}},
\]
with $g$ being the employed prototype filter impulse response (assumed of unit energy) with length $L_{g}$, and $\varphi_{m,n}=(m+n)\frac{\pi}{2}+\phi_{0}$.
Here, it is assumed that $\phi_{0}=mn\pi$. Moreover, usually $L_{g}=KM$, with $K$ being the overlapping factor. 
The corresponding output of a channel with impulse response $h$ of length $L_{h}$ will be 
\begin{equation}
y(l)=\sum_{k=0}^{L_{h}-1}h(k)s(l-k)+w(l),
\label{eq:h}
\end{equation}
with $w(l)$ assumed to be zero mean Gaussian noise with variance $\sigma^{2}$, 
and the analysis filter bank (AFB) output at the $(p,q)$ frequency-time (FT) point is 
\begin{eqnarray}
y_{p,q} & = & \sum_{l}y(l)g_{p,q}^{*}(l) \nonumber \\
& = & \sum_{k=0}^{L_{h}-1}h(k)\sum_{l}s(l-k)g_{p,q}^{*}(l)+\sum_{l}w(l)g_{p,q}^{*}(l) \nonumber \\
& = & \sum_{k=0}^{L_{h}-1}h(k)\sum_{l}\sum_{m=0}^{M-1}\sum_{n}d_{m,n}g\left(l-k-n\frac{M}{2}\right)g\left(l-q\frac{M}{2}\right)\times \nonumber \\
& & e^{j\frac{2\pi}{M}(m-p)\left(l-\frac{L_{g}-1}{2}\right)}e^{-j\frac{2\pi}{M}mk}e^{j(\varphi_{m,n}-\varphi_{p,q})} + \eta_{p,q}\nonumber \\
& = & \sum_{k=0}^{L_{h}-1}h(k)\sum_{m=0}^{M-1}e^{-j\frac{2\pi}{M}mk}\sum_{n}d_{m,n}j^{m+n-p-q}(-1)^{mn-pq}\sum_{l}g\left(l-k-n\frac{M}{2}\right)g\left(l-q\frac{M}{2}\right)\times \nonumber \\
& & e^{j\frac{2\pi}{M}(m-p)\left(l-\frac{L_{g}-1}{2}\right)} + \eta_{p,q}
\label{eq:ypq}
\end{eqnarray}
The latter equation describes the AFB output samples in terms of the channel impulse response, that is, in a \emph{time domain} formulation. Its value is found in the fact that 
it is exact regardless of the relative channel delay spread. Recall that the usual frequency-domain input/output description 
\[
y_{p,q}=H_{p,q}d_{p,q}+\mathrm{interference}+\eta_{p,q}
\]
is only valid (to an approximation) for short enough (relative to the extent of $g$) channels. Eq.~(\ref{eq:ypq}) is thus a good candidate in providing a way to address channel estimation (and equalization) for the general case, namely when the subchannels are not well approximated by the frequency flat model. 

\section{Channel Estimation}
\label{sec:CE}

Consider the preamble-based channel estimation problem. Assume, moreover, that the preamble is constructed in the usual manner, namely consisting of 3~FBMC symbols, with the first and third being guards (all zeros)~\cite{kkrt13}. This is to protect the pilots from being interfered by the unknown data (or control) samples of the previous and current frame.\footnote{Note that the first null FBMC symbol could be also unnecessary, in view of the interframe time gaps commonly used in wireless transmissions. Hence, the preamble as a whole may last from 1 to 1.5 OFDM symbol.}
Then, $d_{m,n}$ is nonzero only for $n=1$. Also, only the case of $q=1$ is of interest. Rewriting eq.~(\ref{eq:ypq}) with these assumptions yields
\begin{eqnarray*}
y_{p,1} & = & \sum_{k=0}^{L_{h}-1}h(k)\sum_{m=0}^{M-1}e^{-j\frac{2\pi}{M}mk}d_{m,1}j^{m-p}(-1)^{m-p}\sum_{l}g\left(l-k-\frac{M}{2}\right)g\left(l-\frac{M}{2}\right)\times \\
& & e^{j\frac{2\pi}{M}(m-p)\left(l-\frac{L_{g}-1}{2}\right)} + \eta_{p,q}
\end{eqnarray*}
or by making use of the time extent of $g$ and making the change of variable $l\leftarrow l-\frac{M}{2}$, 
\begin{eqnarray}
y_{p,1} & = &  \sum_{k=0}^{L_{h}-1}h(k)\underbrace{\sum_{m=0}^{M-1}e^{-j\frac{2\pi}{M}mk}d_{m,1}j^{m-p}e^{-j\frac{2\pi}{M}(m-p)\left(\frac{L_{g}-1}{2}\right)}\sum_{l=k}^{L_{g}-1}g(l-k)g(l)e^{j\frac{2\pi}{M}(m-p)l}}_{\Gamma_{p,k}} + \eta_{p,1} \nonumber \\
& = & \sum_{k=0}^{L_{h}-1}\Gamma_{p,k}h(k)+\eta_{p,1} \label{eq:yp1}
\end{eqnarray}
Note that $\Gamma_{p,k}$ is the response of the entire transmultiplexer to this particular input for a channel equal to a $k$-samples delay, i.e., $h(l)=\delta(l-k)$, and can be easily computed. The AFB output sample is the sum of those responses, multiplied with the channel gains. In matrix-vector notation:
\begin{equation}
\g{y}=\g{\Gamma}\g{h}+\g{\eta},
\label{eq:y=Ah+eta}
\end{equation}
with $\g{y}=\bmx{cccc} y_{0,1} & y_{1,1} & \cdots y_{M-1,1} \emx^{T}$ and similarly for $\g{h}$, while $[\g{\Gamma}]_{p,k}=\Gamma_{p,k}$. 
The matrix $\g{\Gamma}$ is of dimensions $M\times L_{h}$, i.e., tall, and hence the equation above can be solved for the channel impulse response $\g{h}$ using, for example, least squares (LS).
It must be emphasized here that the noise $\eta$ is known to be zero mean Gaussian with the same variance as $w$, however it is not uncorrelated among subcarriers.
Its $M\times M$ covariance matrix is known to be given by 
\begin{equation}
\gss{C}{\sbm{\eta}}{} = \sigma^{2}\bmx{cccccc} 1 & j\beta & 0 & \cdots & 0 & \pm j\beta \\ -j\beta & 1 & j\beta & \cdots & 0 & 0 \\ \vdots & \ddots & \ddots & \ddots & \ddots & \vdots \\ \pm j\beta & 0 & 0 & \cdots & -j\beta & 1 \emx \equiv \sigma^{2}\g{B},
\label{eq:C}
\end{equation}
with $\pm j\beta$ being the correlation of $g_{m,1}$ with $g_{m\pm 1,1}$, \emph{a priori} computable based on the knowledge of $g$~\cite{kkrt13}. Whether it is $j\beta$ or $-j\beta$ at the corners of the matrix $\g{B}$ depends on the employed prototype filter $g$. In the sequel, we will assume $-j\beta$ at the top right corner and $j\beta$ at the bottom left one. Thus, $\g{C}$ is (almost) \emph{tridiagonal} and \emph{circulant} (and hence diagonalizable via DFT). For more details, see~\cite{kkrt10}. The invertibility of this matrix is investigated in~\cite{kbrhk11} and can be seen to be ensured for all practical purposes. 

In view of the color of the noise component in~\pref{eq:y=Ah+eta}, \emph{weighted LS} (i.e., Gauss-Markov) estimation should be used instead, resulting in the following estimate for $\g{h}$:
\begin{equation}
\g{h}=(\gss{\Gamma}{}{H}\gss{C}{\sbm{\eta}}{-1}\g{\Gamma})^{-1}\gss{\Gamma}{}{H}\gss{C}{\sbm{\eta}}{-1}\g{y}
\label{eq:hest}
\end{equation}
Because of the Gaussianity of the noise, this can be seen to be also the maximum likelihood (ML) estimate. 

Of course, for a sparse preamble, where $d_{m,1}\neq 0$ only for some isolated subcarriers $m\in \mathcal{P}$, the above holds with the sum for $m$ being replaced by a sum on $\mathcal{P}$. 

\section{Optimal Preamble Design}
\label{sec:optimal}

Rewrite $\Gamma_{p,k}$ as
\begin{eqnarray*}
\Gamma_{p,k} & = & \sum_{m=0}^{M-1}d_{m,1}\underbrace{j^{m-p}e^{-j\frac{2\pi}{M}mk}e^{-j\frac{2\pi}{M}(m-p)\left(\frac{L_{g}-1}{2}\right)}\sum_{l=k}^{L_{g}-1}g(l-k)g(l)e^{j\frac{2\pi}{M}(m-p)l}}_{\mathcal{G}_{p,k}^{*}(m)} \\
& = & \gss{\mathcal{G}}{p,k}{H}\g{d},
\end{eqnarray*}
with the obvious definition for the $M\times 1$ vectors $\gss{\mathcal{G}}{p,k}{}$ and $\g{d}$. 
One can then express the matrix $\g{\Gamma}$ in the form:
\begin{equation}
\g{\Gamma}=\g{\mathcal{G}}\g{D},
\label{eq:A=GD}
\end{equation}
where 
\begin{eqnarray*}
\g{\mathcal{G}} & = & 
\bmx{cccc} 
\gss{\mathcal{G}}{0,0}{H} & \gss{\mathcal{G}}{0,1}{H} & \cdots & \gss{\mathcal{G}}{0,L_{h}-1}{H} \\
\gss{\mathcal{G}}{1,0}{H} & \gss{\mathcal{G}}{1,1}{H} & \cdots & \gss{\mathcal{G}}{1,L_{h}-1}{H} \\
\vdots & \vdots & \ddots & \vdots \\
\gss{\mathcal{G}}{M-1,0}{H} & \gss{\mathcal{G}}{M-1,1}{H} & \cdots & \gss{\mathcal{G}}{M-1,L_{h}-1}{H}
\emx \mbox{\ \ \ } (M\times ML_{h})
\end{eqnarray*}
and
\[
\g{D} = \gss{I}{L_{h}}{}\otimes \g{d}, \mbox{\ \ \ } (ML_{h}\times L_{h})
\]
with $\otimes$ denoting the (left) Kronecker product. 
The matrix $\g{\mathcal{G}}$ is normally banded, in view of the limited overlapping of the $g_{m,n}$ functions in time and frequency. More on its structure needs to be 
investigated, based also on the results of the appendix in~\cite{kkrt13}. See next subsection for an elaboration on the structure of $\g{\mathcal{G}}$. 

Optimal design of the preamble $d_{m,1}$, $m=0,1,\ldots,M-1$, could be stated with the aid of the above expression for $\g{\Gamma}$ as follows:
\begin{equation}
\min_{\sbm{d}}\tr\left\{(\gss{\Gamma}{}{H}\gss{C}{\sbm{\eta}}{-1}\g{\Gamma})^{-1}\right\}
\label{eq:MMSE}
\end{equation} 
subject to a constraint on the transmitted energy. The latter could refer to the energy input to the SFB (e.g., $\sum_{m=0}^{M-1}|d_{m,1}|^{2}$) or, more realistically, to the energy of the
modulated preamble at the SFB output as in~\cite{kkrt10,kbrhk11}. For a full preamble, the latter is not trivially related to the former.
Regarding the optimality criterion above, it is known that $\g{d}$ will be optimal if it diagonalizes the matrix $(\gss{\Gamma}{}{H}\gss{C}{\sbm{\eta}}{-1}\g{\Gamma})^{-1}$ while satisfying the energy constraint. 

\subsection{Full Preamble}
\label{sec:optimal_full}

In view of the fact that a given subcarrier only overlaps with the immediately adjacent ones, it should be expected that no more than three of the entries of each of the vectors $\gss{\mathcal{G}}{p,k}{}$ is nonzero. Indeed, this can be shown, and in fact the $\gss{\mathcal{G}}{p,0}{}$ vectors can be easily written in terms of the matrix $\g{B}$ defined above.
In general,
\begin{equation}
\g{\mathcal{G}} = \bmx{ccccccc} \gss{\mathcal{G}}{0}{} & \vdots & \gss{\mathcal{G}}{1}{} & \vdots & \cdots & \vdots & \gss{\mathcal{G}}{L_{h}-1}{} \emx,
\label{eq:Gcal}
\end{equation}
where, for $k=0,1,\ldots,L_{h}-1$, 
\[
\gss{\mathcal{G}}{k}{}=\bmx{cccc} \gss{\mathcal{G}}{0,k}{} & \gss{\mathcal{G}}{1,k}{} & \cdots & \gss{\mathcal{G}}{M-1,k}{} \emx^{H}
\]
is an $M\times M$ matrix. One can readily verify that 
\begin{equation}
\gss{\mathcal{G}}{0}{} = \g{B}.
\label{eq:G0=B}
\end{equation}
Recall that $\gss{\mathcal{G}}{k}{}\g{d}$ is the response of the entire system when the channel is only a delay of $k$ samples. Hence, $\gss{\mathcal{G}}{0}{}\g{d}$ is the vector of pseudo-pilots, already known to equal $\g{B}\g{d}$. 
The rest of the blocks of $\g{\mathcal{G}}$ can also be computed \emph{a priori} and shown to be tridiagonal. Thus, at every row of $\g{\Gamma}$, only three of the pilots appear, but they appear $L_{h}$ times. The vector $\g{d}$ will be assumed to be complex-valued in the sequel, as the optimal preamble does not have to be OQAM modulated (see, e.g., the IAM-I and (E)IAM-C preambles in~\cite{kkrt13}).

In view of the above, the cost function in~(\ref{eq:MMSE}) becomes:
\begin{equation}
\mathrm{MSE} = \sigma^{2}\tr\left\{\bmx{cccc} \gss{d}{}{H}\g{B}\g{d} & \gss{d}{}{H}\gss{\mathcal{G}}{1}{}\g{d} & \cdots & \gss{d}{}{H}\gss{\mathcal{G}}{L_{h}-1}{}\g{d} \\ \gss{d}{}{H}\gss{\mathcal{G}}{1}{H}\g{d} & \gss{d}{}{H}\gss{\mathcal{G}}{1}{H}\gss{B}{}{-1}\gss{\mathcal{G}}{1}{}\g{d} & \cdots & \gss{d}{}{H}\gss{\mathcal{G}}{1}{H}\gss{B}{}{-1}\gss{\mathcal{G}}{L_{h}-1}{}\g{d} \\ \vdots & \vdots & \ddots & \vdots \\ 
\gss{d}{}{H}\gss{\mathcal{G}}{L_{h}-1}{H}\g{d} & \gss{d}{}{H}\gss{\mathcal{G}}{L_{h}-1}{H}\gss{B}{}{-1}\gss{\mathcal{G}}{1}{}\g{d} & \cdots & \gss{d}{}{H}\gss{\mathcal{G}}{L_{h}-1}{H}\gss{B}{}{-1}\gss{\mathcal{G}}{L_{h}-1}{}\g{d} \emx^{-1}\right\},
\label{eq:MSE}
\end{equation}
where the Hermitian symmetry of $\g{B}$ was taken into account. The optimal preamble design problem can thus be stated as the minimization of~(\ref{eq:MSE}) with respect to the vector $\g{d}$ subject to a constraint on the SFB output energy, say 
\begin{equation}
\gss{d}{}{H}\g{B}\g{d} \leq \mathcal{E}
\label{eq:aBa}
\end{equation}
It is known that the above problem is directly connected to the diagonalization (via the proper choice of $\g{d}$) of the matrix involved in the MSE expression. Thus, the following system of $\frac{L_{h}(L_{h}-1)}{2}$ (quadratic) equations in $\g{d}$ results:
\begin{eqnarray}
\gss{d}{}{H}\gss{\mathcal{G}}{k}{}\g{d} & = & 0, \mbox{\ \ \ } k=1,2,\ldots,L_{h}-1 \label{eq:ortho1}\\
\gss{d}{}{H}\gss{\mathcal{G}}{k}{H}\gss{B}{}{-1}\gss{\mathcal{G}}{l}{}\g{d} & = & 0, \mbox{\ \ \ } k=1,2,\ldots,L_{h}-2, l>k \label{eq:ortho2}
\end{eqnarray}

In view of the circulant structure of the matrix $\g{B}$, the above can be alternatively written in a form which may be more convenient for deducing the solution for $\g{d}$. Indeed, $\g{B}$ is diagonalized with the DFT matrix, that is,
\begin{equation}
\g{B}=\g{F}\g{\Lambda}\gss{F}{}{H},
\label{eq:B=FLF}
\end{equation}
where $\g{F}$ is the \emph{unitary} $M$-point DFT matrix and $\g{\Lambda}$ is diagonal with the (real and positive) eigenvalues of $\g{B}$ on its main diagonal. (For a closed form expression of these eigenvalues, see~\cite{kbrhk11}.)
Recalling that the noise term in~\pref{eq:y=Ah+eta} has covariance given by~\pref{eq:C}, the premultiplication of~\pref{eq:y=Ah+eta} by a square root of $\g{B}$ dictated by~\pref{eq:B=FLF}, namely $\gss{\Lambda}{}{-1/2}\gss{F}{}{H}$, whitens the noise and leads to the simpler input/output equation:
\[
\g{\tilde{y}}=\bmx{ccccccc} \gss{\Lambda}{}{1/2}\gss{F}{}{H} & \vdots & \gss{\Lambda}{}{-1/2}\gss{F}{}{H}\gss{\mathcal{G}}{1}{} & \vdots & \cdots & \vdots &  \gss{\Lambda}{}{-1/2}\gss{F}{}{H}\gss{\mathcal{G}}{L_{h}-1}{}\emx \g{D}\g{h}+\g{\tilde{\eta}},
\]
where $\g{\tilde{y}},\g{\tilde{\eta}}$ denote the transformed variables (with $\g{\tilde{\eta}}$ being $\g{\mathcal{CN}}(\g{0},\sigma^{2}\gss{I}{M}{})$) and use was made of eq.~\pref{eq:G0=B}. This can be further written as
\begin{eqnarray}
\g{\tilde{y}} & = & \underbrace{\bmx{cccc} \gss{I}{M}{} & \gss{\tilde{\mathcal{G}}}{1}{} & \cdots & \gss{\tilde{\mathcal{G}}}{L_{h}-1}{} \emx}_{\g{\tilde{\mathcal{G}}}} \g{\tilde{D}}\g{h}+\g{\tilde{\eta}} \nonumber \\
& = & \g{\tilde{\mathcal{G}}}\g{\tilde{D}}\g{h}+\g{\tilde{\eta}} \label{eq:yt=Gtath+etat} \\
& = & \g{\tilde{\Gamma}}\g{h}+\g{\tilde{\eta}}, \label{eq:yt=Ath+etat}
\end{eqnarray}
where
\begin{eqnarray}
\gss{\tilde{\mathcal{G}}}{k}{} & = & \gss{\Lambda}{}{-1/2}\gss{F}{}{H}\gss{\mathcal{G}}{k}{}\g{F}\gss{\Lambda}{}{-1/2}, \label{eq:Gcaltildek}\\
\g{\tilde{D}} & = & \gss{I}{L_{h}}{} \otimes \gss{\Lambda}{}{1/2}\gss{F}{}{H}\g{d} 
\end{eqnarray}
Note that $\g{\tilde{d}}=\gss{\Lambda}{}{1/2}\gss{F}{}{H}\g{d}$ is a weighted version of the output of the IDFT block in the polyphase realization of the synthesis filter bank, and hence the input to the polyphase network~\cite{ssl02}. 
Moreover, $\g{\tilde{y}}$ is closely related to the output of the polyphase network at the analysis filter bank. 

Thus, the optimality criterion is now expressed as 
\begin{eqnarray}
& & \min_{\sbm{\tilde{d}}}\sigma^{2}\mathrm{tr}\left\{(\gss{\tilde{\Gamma}}{}{H}\g{\tilde{\Gamma}})^{-1}\right\} \label{eq:newMMSE} \\
& \mathrm{s.t.} & \|\g{\tilde{d}}\|^{2}\leq \mathcal{E} \label{eq:aLa}
\end{eqnarray}
As previously, the above amounts to the requirement that $\gss{\tilde{\Gamma}}{}{H}\g{\tilde{\Gamma}}$ is diagonal, that is, that the columns of the matrix $\g{\tilde{\Gamma}}$ are orthogonal. 
To aid the analysis of this matrix, the structure of the blocks of~\pref{eq:Gcal} will be studied first. It can be shown (details are omitted here) that it is not only $\gss{\mathcal{G}}{0}{}$ that has a circulant structure. The rest of the blocks also contain circulant matrices. Specifically:
\begin{equation}
\gss{\mathcal{G}}{k}{} = \gss{W}{}{k}\gss{G}{k}{}, \mbox{\ \ \ } k=0,1,2,\ldots,L_{h}-1
\label{eq:Gcalk}
\end{equation}
where 
\[
\g{W} = \diag(1,e^{-j\frac{2\pi}{M}},e^{-j2\frac{2\pi}{M}},\ldots,e^{-j(M-1)\frac{2\pi}{M}})
\]
and $\gss{G}{k}{}$ is a Hermitian \emph{tridiagonal} circulant matrix. (Details about its entries are omitted.)
It is readily verified that\footnote{It is of interest to note that this identity states the well known duality between frequency- (expressed by $\g{W}$) and time- (expressed by $\g{Z}$) shifts. By the way, it is shown in~\cite{mbh13} that these two matrices, $\g{W}$ and $\g{Z}$, play a fundamental role in describing a channel in the time-frequency domain. Notably, small frequency and time shifts commute approximately (i.e., $\gss{W}{}{k}\gss{Z}{}{l}\approx \gss{Z}{}{l}\gss{W}{}{k}$ for $k,l\ll M$), a property that underlies the behavior of \emph{underspread} channels~\cite{mbh13}.} 
\begin{equation}
\gss{F}{}{H}\gss{W}{}{k}=\gss{Z}{}{k}\gss{F}{}{H},
\label{eq:FW=ZF}
\end{equation}
where $\gss{Z}{}{k}$ is the $M\times M$ permutation matrix that, when premultiplying a matrix, circularly shifts its rows downwards by $k$. That is,
\[
\g{Z}=\bmx{cc} \gss{0}{1\times (M-1)}{} & 1 \\ \gss{I}{M-1}{} & \gss{0}{(M-1)\times 1}{} \emx
\]
Hence~\pref{eq:Gcaltildek} becomes
\begin{eqnarray}
\gss{\tilde{\mathcal{G}}}{k}{} & = &  \gss{\Lambda}{}{-1/2}\gss{Z}{}{k}\gss{F}{}{H}\gss{G}{k}{}\g{F}\gss{\Lambda}{}{-1/2} \nonumber \\
& = & \gss{\Lambda}{}{-1/2}\gss{Z}{}{k}\gss{\Lambda}{k}{}\gss{\Lambda}{}{-1/2},
\label{eq:Gcaltildek=diagonal}
\end{eqnarray}
with $\gss{\Lambda}{k}{}$ being diagonal with the eigenvalues of $\gss{G}{k}{}$ on its main diagonal. Clearly, these are real valued. Note that $\gss{\Lambda}{0}{}=\g{\Lambda}$ and $\gss{Z}{}{0}=\gss{I}{M}{}$. 
Thus, one can write the $\g{\tilde{\mathcal{G}}}$ matrix as
\begin{equation}
\g{\tilde{\mathcal{G}}}=\bmx{ccccc} \gss{I}{M}{} & \gss{\Lambda}{}{-1/2}\g{Z}\gss{\Lambda}{1}{}\gss{\Lambda}{}{-1/2} & \gss{\Lambda}{}{-1/2}\gss{Z}{}{2}\gss{\Lambda}{2}{}\gss{\Lambda}{}{-1/2} & \cdots & \gss{\Lambda}{}{-1/2}\gss{Z}{}{L_{h}-1}\gss{\Lambda}{L_{h}-1}{}\gss{\Lambda}{}{-1/2} \emx
\label{eq:Gcaltilde}
\end{equation}
and hence the $\g{\tilde{\Gamma}}$ matrix becomes
\begin{equation}
\g{\tilde{\Gamma}} = \bmx{ccccc} \g{\tilde{d}} & \gss{\Lambda}{}{-1/2}\g{Z}\gss{\Lambda}{1}{}\gss{\Lambda}{}{-1/2}\g{\tilde{d}} & \gss{\Lambda}{}{-1/2}\gss{Z}{}{2}\gss{\Lambda}{2}{}\gss{\Lambda}{}{-1/2}\g{\tilde{d}} & \cdots & \gss{\Lambda}{}{-1/2}\gss{Z}{}{L_{h}-1}\gss{\Lambda}{L_{h}-1}{}\gss{\Lambda}{}{-1/2}\g{\tilde{d}} \emx 
\label{eq:Atilde}
\end{equation}
It is this matrix that is desired to have orthogonal columns. 
The orthogonality of the columns of the above matrix translates now to the $\frac{L_{h}(L_{h}-1)}{2}$ requirements
\begin{equation}
\gss{\tilde{d}}{}{H}\gss{\Lambda}{}{-1/2}\gss{\Lambda}{k}{}(\gss{Z}{}{k})^{T}\gss{\Lambda}{}{-1}\gss{Z}{}{l}\gss{\Lambda}{l}{}\gss{\Lambda}{}{-1/2}\g{\tilde{d}}=0, \mbox{\ \ for\ } k\neq l,
\label{eq:ortho}
\end{equation}
where it should be reminded that $\gss{Z}{}{0}=\gss{Z}{}{M}=\gss{I}{M}{}$ and $\gss{\Lambda}{0}{}=\g{\Lambda}$. 
Observe that 
\[
(\gss{Z}{}{k})^{T}=\bmx{cc} \gss{0}{(M-k)\times k}{} & \gss{I}{M-k}{} \\ \gss{I}{k}{} & \gss{0}{k\times (M-k)}{} \emx=\gss{Z}{}{M-k}
\]
Then, it can be verified that the matrix in~\pref{eq:ortho} becomes (for $k\geq l$) 
\begin{equation}
\gss{\Lambda}{}{-1/2}\gss{\Lambda}{k}{}(\gss{Z}{}{k})^{T}\gss{\Lambda}{}{-1}\gss{Z}{}{l}\gss{\Lambda}{l}{}\gss{\Lambda}{}{-1/2} =
\bmx{ccc} \gss{0}{(M-k)\times (k-l)}{} & \gss{L}{M-k}{} & \gss{0}{(M-k)\times l}{} \\ \gss{0}{l\times (k-l)}{} & \gss{0}{l\times (M-k)}{} & \gss{L}{l}{} \\
\gss{L}{k-l}{} & \gss{0}{(k-l)\times (M-k)}{} & \gss{0}{(k-l)\times l}{} \emx
\label{eq:k>l}
\end{equation}
with $\gss{L}{s}{}$ being a diagonal matrix of order $s$ constructed from the entries $\lambda_{k,\cdot},\lambda_{l,\cdot},\lambda_{\cdot}$ of $\gss{\Lambda}{k}{},\gss{\Lambda}{l}{},\g{\Lambda}$, respectively, as follows 
\begin{eqnarray}
\gssa{L}{M-k}{}{i,i} & = & \frac{\rho_{k,i}\rho_{l,k-l+i}}{\lambda_{k+i}}, \mbox{\ \ \ } i=1,2,\ldots,M-k, \\
\gssa{L}{l}{}{i,i} & = & \frac{\rho_{k,M-k+i}\rho_{l,M-l+i}}{\lambda_{i}}, \mbox{\ \ \ } i=1,2,\ldots,l, \\
\gssa{L}{k-l}{}{i,i} & = & \frac{\rho_{k,M-k+l+i}\rho_{l,i}}{\lambda_{l+i}}, \mbox{\ \ \ } i=1,2,\ldots,k-l,
\end{eqnarray}
where, for notational convenience, $\rho_{k,i}=\frac{\lambda_{k,i}}{\lambda_{i}^{1/2}}$ and similarly for $\rho_{l,i}$. The orthogonality conditions thus are given by~\pref{eq:ortho} and~\pref{eq:k>l} for $k>l$. When $k=l$, \pref{eq:k>l} becomes
\[
\gss{\Lambda}{}{-1/2}\gss{\Lambda}{k}{}(\gss{Z}{}{k})^{T}\gss{\Lambda}{}{-1}\gss{Z}{}{k}\gss{\Lambda}{k}{}\gss{\Lambda}{}{-1/2}=
\bmx{cc} \gss{L}{M-k}{} & \gss{0}{(M-k)\times k}{} \\ \gss{0}{k\times (M-k)}{} & \gss{L}{k}{} \emx,
\]
that is, diagonal, with the (positive) diagonal entries
\begin{eqnarray}
\gssa{L}{M-k}{}{i,i} & = & \frac{\rho_{k,i}^{2}}{\lambda_{k+i}}=\frac{\lambda_{k,i}^{2}}{\lambda_{i}\lambda_{k+i}}, \mbox{\ \ \ } i=1,2,\ldots,M-k, \label{eq:L1} \\
\gssa{L}{k}{}{i,i} & = & \frac{\rho_{k,M-k+i}^{2}}{\lambda_{i}}=\frac{\lambda_{k,M-k+i}^{2}}{\lambda_{M-k+i}\lambda_{i}}, \mbox{\ \ \ } i=1,2,\ldots,k \label{eq:L2}
\end{eqnarray}
Observe that these are the ratios of the squared diagonal entries of $\gss{\Lambda}{k}{}$ over the diagonal entries of $\g{\Lambda}$ and the diagonal entries of $\g{\Lambda}$ circularly shifted upwards by $k$, that is, $\gss{Z}{}{M-k}\g{\Lambda}$. 
Hence, defining the diagonal matrix 
\[
\gss{\Delta}{k}{}=\gss{\Lambda}{}{-1}\gss{\Lambda}{k}{2}\left(\gss{Z}{}{M-k}\g{\Lambda}\right)^{-1},
\]
the diagonal entries of $\gss{\tilde{\Gamma}}{}{H}\g{\tilde{\Gamma}}$ are given by
\begin{equation}
\delta_{k}=\gss{\tilde{d}}{}{H}\gss{\Delta}{k}{}\g{\tilde{d}}
\label{eq:delta_k}
\end{equation}
and are clearly nonnegative. 
A particular case is for $k=l=0$, where the above matrix reduces to the identity, as it should. Then the latter quantity equals $\|\g{\tilde{d}}\|^{2}$. 
Assuming that $\g{\tilde{d}}$ was so chosen as to make the columns of $\g{\tilde{\Gamma}}$ orthogonal, the trace in~\pref{eq:newMMSE} can be written as
\begin{equation}
\mathrm{tr}\left\{(\gss{\tilde{\Gamma}}{}{H}\g{\tilde{\Gamma}})^{-1}\right\}=\sum_{k=0}^{L_{h}-1}\frac{1}{\delta_{k}}=\frac{1}{\|\g{\tilde{d}}\|^{2}}+\sum_{k=1}^{L_{h}-1}\frac{1}{\delta_{k}},
\label{eq:mintrace}
\end{equation}
implying that $\delta_{k}$, $k=1,2,\ldots,L_{h}-1$ must be maximized, with $\delta_{0}=\|\g{\tilde{d}}\|^{2}$ not exceeding $\mathcal{E}$. 

From~\pref{eq:delta_k} and~\pref{eq:aLa}, it follows that $\delta_{k}$ is maximized for $\g{\tilde{d}}$ equal to an eigenvector of $\gss{\Delta}{k}{}$ associated to its largest eigenvalue.
Since $\gss{\Delta}{k}{}$ is diagonal, this means that $\g{\tilde{d}}$ should be (proportional to) the vector with all zeros except at the positions corresponding to the largest diagonal entries of $\gss{\Delta}{k}{}$. One can readily see that such $\g{\tilde{d}}$ vectors would also satisfy~\pref{eq:ortho}, provided their nonzero entries are located sufficiently far apart. In view of the relation with $\g{d}$, the latter would then equal a linear combination of the corresponding columns of the matrix $\g{F}$, with the coefficients specified by $\mathcal{E}$ and the corresponding $\lambda$'s. 

However, the maximum entries of the $\gss{\Delta}{k}{}$'s are not found at the same positions in all these matrices. Hence it is not obvious what will be the solution for $\g{\tilde{d}}$ that maximizes \emph{each} of the $\delta_{k}$'s. A simple answer to that problem is to chose $\g{\tilde{d}}$ as
\begin{equation}
\g{\tilde{d}}=\sqrt{\mathcal{E}}\gss{e}{m}{},
\label{eq:atilde}
\end{equation}
where $\gss{e}{m}{}$ denotes the $m$th pinning vector, namely a vector of all zeros except for a unity at the $m$th position, and with $m=1,2,\ldots,M$ to be determined. In fact, such would be the maximizing vector for $\delta_{k}$ if the maximum diagonal entry of $\gss{\Delta}{k}{}$ were unique (and located at the $m$th position on the main diagonal). Any $m$ can do, in terms of satisfying~\pref{eq:ortho}. To find the ``optimum'' $m$, the one that minimizes the value of the trace above, write~\pref{eq:mintrace} for $\g{\tilde{d}}=\sqrt{\mathcal{E}}\gss{e}{m}{}$. The result is
\[
\mathrm{tr}\left\{(\gss{\tilde{\Gamma}}{}{H}\g{\tilde{\Gamma}})^{-1}\right\}=\frac{1}{\mathcal{E}}\left[1+\lambda_{m}\sum_{k=1}^{L_{h}-1}\frac{\lambda_{((k+m))_{M}}}{\lambda_{k,m}^{2}}\right]
\]
where the notation $((\cdot))_{M}$ is used here to denote that the argument is circularly shifted back to the set $\{1,2,\ldots,M\}$ if it exceeds $M$. For example, $((M+1))_{M}=1,((M+5))_{M}=5,((2))_{M}=2$. Then, choose $m$ as 
\begin{equation}
m_{\mathrm{opt}} = \arg\min_{m}\lambda_{m}\sum_{k=1}^{L_{h}-1}\frac{\lambda_{((k+m))_{M}}}{\lambda_{k,m}^{2}}
\label{eq:mopt}
\end{equation}
Note that this choice for $\g{\tilde{d}}$ is only optimal among the vectors with only one nonzero entry and it yields an $\g{d}$ vector equal to
\begin{equation}
\g{d}=\sqrt{\frac{\mathcal{E}}{\lambda_{m_{\mathrm{opt}}}}}\gss{f}{m_{\mathrm{opt}}}{},
\label{eq:aopt_singleton}
\end{equation}
where $\gss{f}{m}{}$ is the $m$th column of the matrix $\g{F}$. 

\medskip

\noindent
\emph{Remarks.}
\begin{enumerate}
\item As shown in the appendix, when $L_{h}$ is quite small compared to $M$, and hence the assumption of flat subchannels is accurate, $\gss{\Lambda}{k}{}\approx\g{\Lambda}$ for all $k$. This implies that the MSE can be written as
\[
\mathrm{MSE}\approx\frac{\sigma^{2}}{\mathcal{E}}\left[1+\frac{1}{\lambda_{m}}\sum_{k=1}^{L_{h}-1}\lambda_{((k+m))_{M}}\right]
\] 
It thus seems that $m_{\mathrm{opt}}$ should be the index of the largest eigenvalue of $\g{B}$. The solution in that case would be given by~\pref{eq:atilde}. In the appendix, it will be shown (via a different approach) that this is indeed the optimal solution for this case (confirming also the solution derived in~\cite{kbrhk11}). 
With $L_{h}<<M$, it may be reasonable to expect that $\lambda_{((k+m))_{M}}\approx \lambda_{m}$ for $k=1,2,\ldots,L_{h}-1$. Then
\[
\mathrm{MSE}\approx\frac{\sigma^{2}}{\mathcal{E}}L_{h}
\]
This is the MSE achieved with optimal sparse preamble (see the appendix).
\item The above cost function is the MSE for impulse response estimation, what is sometimes called time-domain MSE~\cite{kkrt10}. An estimate for the channel frequency response (CFR) can then be computed as
\[
\g{\hat{H}}=\sqrt{M}\ga{F}{:,1:L_{h}}\g{\hat{h}},
\]
where $\ga{F}{:,1:L_{h}}$ denotes the $M\times L_{h}$ matrix consisting of the first $L_{h}$ columns of $\g{F}$, employing Matlab notation. The frequency-domain MSE will then be given by
\begin{eqnarray*}
\mathrm{tr}\left\{\gss{C}{\sbm{\hat{H}}}{}\right\} & = & M\mathrm{tr}\left\{\ga{F}{:,1:L_{h}}\gss{C}{\sbm{\hat{h}}}{}\gssa{F}{}{H}{:,1:L_{h}}\right\} \\
& = & M\mathrm{tr}\left\{\gss{C}{\sbm{\hat{h}}}{}\gssa{F}{}{H}{:,1:L_{h}}\ga{F}{:,1:L_{h}}\right\}
\end{eqnarray*}
where $\gssa{F}{}{H}{:,1:L_{h}}\ga{F}{:,1:L_{h}}=\gss{I}{L_{h}}{}$. Thus, with the CFR estimated as above, the MSEs are related as 
\begin{equation}
\mathrm{tr}\left\{\gss{C}{\sbm{\hat{H}}}{}\right\}=M\mathrm{tr}\left\{\gss{C}{\sbm{\hat{h}}}{}\right\}
\label{eq:MSEf=M*MSEt}
\end{equation}
\item The estimator in~\pref{eq:hest} can be equivalently written as
\[
\g{\hat{h}}=\left(\gss{\tilde{\Gamma}}{}{H}\g{\tilde{\Gamma}}\right)^{-1}\gss{\tilde{\Gamma}}{}{H}\g{\tilde{y}}
\]
With $\g{\tilde{d}}$ chosen as above, the factor $\left(\gss{\tilde{\Gamma}}{}{H}\g{\tilde{\Gamma}}\right)^{-1}$ is diagonal. This simplifies considerably the computation of the estimator matrix, which is anyway performed offline. The online computation of $\g{\hat{h}}$ requires $ML_{h}$ complex multiplications. 
\end{enumerate}

\begin{figure}
\begin{center}
\begin{tabular}{c}
\includegraphics[width=12cm]{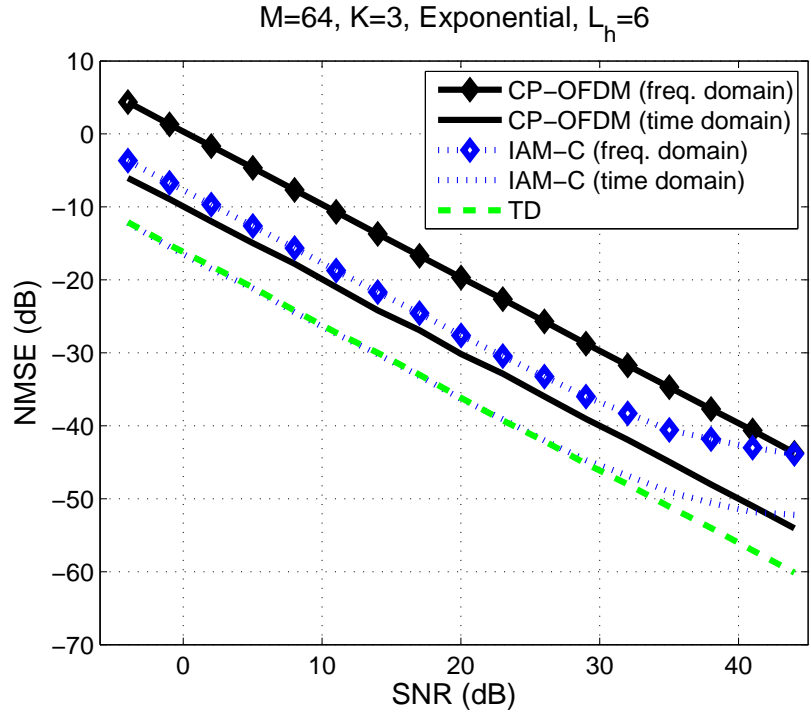} \\
(a) \\
 \includegraphics[width=12cm]{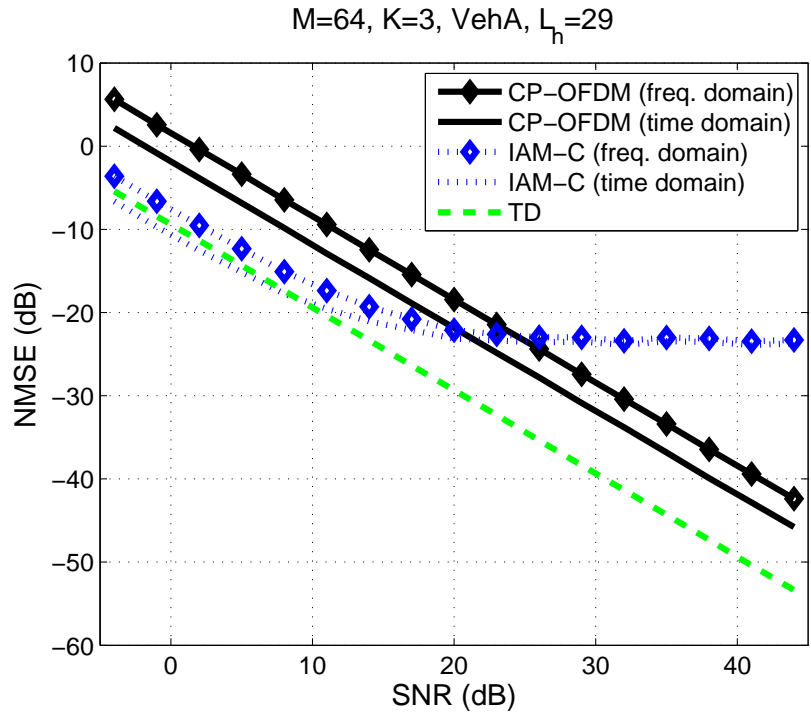} \\
 (b)
 \end{tabular}
 \end{center}
  \caption{MSE performance of the methods under study, with optimal full preambles, for channels of (a) low and (b) high frequency selectivity.}\label{fig:MSEfull}
\end{figure}

Training with full preambles designed as above (\pref{eq:mopt}, \pref{eq:aopt_singleton}) leads to the results shown in Fig.~\ref{fig:MSEfull}. 
Filter banks designed as in~\cite{b01} were employed, with $M=64$ and $K=3$. 
Two examples corresponding to channels with relatively low and high frequency selectivity are provided. The frequency domain normalized MSE (NMSE) is plotted, that is $E\left(\frac{\|\g{H}-\g{\hat{H}}\|^{2}}{\|\g{H}\|^{2}}\right)$. Due to its optimality among the preambles of the given structure for channels of low delay spread, the IAM-C method is employed as a benchmark representative of the methods assuming flat subchannels~\cite{kkrt13}. CP-OFDM is also included in the comparison. The CP used was the shortest allowed, namely equal to the channel order. For CP-OFDM, the full preamble is optimally chosen as one of the columns of the DFT matrix, as suggested by~\cite{kkrt10}. For IAM-C, the preamble is as in~\cite{kbrhk11}. For a fair comparison, the SFB output training signals for the FBMC/OQAM methods were scaled so as to have the same power with that of CP-OFDM. 
For the sake of the completeness of the comparison, and to demonstrate the effect of the channel length knowledge, the frequency-domain methods (i.e., IAM-C and CP-OFDM) are also enriched with a DFT-interpolation step. 
This consists of filtering the CFR estimates through taking the inverse DFT and truncating it to the assumed known length $L_{h}$ (see~\cite{cd02,lcych12,krbh13}). To simplify these steps, sampled-spaced channels will be assumed in the sequel~\cite{esbwb00}. 
These variations will be henceforth called ``time domain'', to distinguish them from the ``frequency domain'' estimation methods. The method developed above (henceforth referred to as \emph{TD} (time domain)) outperforms IAM-C in both cases and as expected does not exhibit the well known error floors at high SNR values. Nevertheless, in the highly frequency selective case (Fig.~\ref{fig:MSEfull}(b)), it is slightly worse than the time domain IAM at low to moderate SNRs. This may be due to the inadequateness of a single FBMC preamble symbol in this case. Analogous results will be obtained for sparse preambles shortly. Extending the method to employ longer preambles seems to be a necessity in overcoming such difficulties and improving the estimation performance (see Section~\ref{sec:longer}). 
Nevertheless, it must be emphasized that this performance is achieved with only one pilot FBMC symbol, in contrast to other methods (e.g., \cite{sir08,W09,PHYDYAS_D3,bnn12,nbn13}) requiring long training sequences that imply considerable bandwidth loss, complexity increase, and lack of robustness to channel time variations. 

\medskip

\noindent
\emph{Remark.}
Note that the improvement from following a time-domain approach over a frequency domain approach is diminishing as the channel length increases with respect to $M$.
This should be expected in view of results reported in~\cite{esbwb00,cd02} on the MSEs for CFR estimation achieved through time-domain and frequency domain approaches in CP-OFDM.\footnote{In fact, what we call here ``time domain'' approach is the estimator~C studied in~\cite{esbwb00}.} Namely, the time domain approach is better by a factor of $\frac{L_{h}}{M}$ (i.e., MSE lower by $10\log_{10}\frac{M}{L_{h}}$~dB).\footnote{More generally, the improvement factor is $\frac{L}{M}$, if $L\geq L_{h}$ nonnegligible taps were considered for the channel. Here, $L=L_{h}$.} This can be verified in the above figures. The two approaches result in the same MSE at the extreme case $L_{h}=M$. Similar results are seen to hold for IAM-C. Specifically, as shown in~\cite{kkrt10}, the DFT interpolation (i.e., projection onto the space of $L_{h}$-long channels), referred to as ``time domain'' approach, improves the MSE of IAM by $10\log_{10}\left(\frac{M}{L_{h}}\frac{1}{1+2\beta}\right)$~dB. This improvement can be observed in the figures, particularly in the low frequency selective channel case (a) (where the assumptions leading to the results of~\cite{kkrt10} hold true). 

\subsection{Sparse Preamble}
\label{sec:optimal_sparse}

In a sparse preamble, only the symbols loaded on a set of isolated (surrounded by nulls) subcarriers are nonzeros. Clearly, to estimate $L_{h}$ parameters, at least an equal number of equations are needed. Henceforth, it will be assumed that there are no more than $L_{h}$ pilot tones, indexed as $\mathcal{P}=\{p_{1},p_{2},\ldots,p_{L_{h}}\}\subset \{0,1,\ldots,M-1\}$. Let $\gss{d}{\mathcal{P}}{}=\bmx{cccc} d_{p_{1}} & d_{p_{2}} & \cdots & d_{p_{L_{h}}} \emx^{T}$ be the vector of pilots loaded on the pilot subcarriers. Then~\pref{eq:A=GD} holds as
\begin{equation}
\gss{\Gamma}{\mathcal{P}}{}=\gss{\mathcal{G}}{\mathcal{P}}{}\gss{D}{\mathcal{P}}{},
\label{eq:AP=GPDP}
\end{equation}
where 
\[
\gss{D}{\mathcal{P}}{}=\gss{I}{L_{h}}{}\otimes \gss{d}{\mathcal{P}}{}
\]
and
\[
\gss{\mathcal{G}}{\mathcal{P}}{}=\bmx{ccccc} \gss{\mathcal{G}}{0|\mathcal{P}}{} & \gss{\mathcal{G}}{1|\mathcal{P}}{} & \gss{\mathcal{G}}{2|\mathcal{P}}{} & \cdots & \gss{\mathcal{G}}{L_{h}-1|\mathcal{P}}{} \emx
\]
with $\gss{\mathcal{G}}{k|\mathcal{P}}{}$ denoting the $L_{h}\times L_{h}$ submatrix of $\gss{\mathcal{G}}{k}{}$ consisting of its rows and columns indexed by $\mathcal{P}$. 
Similarly,
\[
\gss{\mathcal{G}}{k|\mathcal{P}}{}=\gss{W}{\mathcal{P}}{k}\gss{G}{k|\mathcal{P}}{}
\]
for $k=0,1,\ldots,L_{h}-1$, with obvious notation. What is very interesting to observe is that, \emph{regardless} of the choice of $\mathcal{P}$ (and provided that there are no consecutive pilot tones), the structure of the $\gss{G}{k}{}$ matrices implies that 
\begin{equation}
\gss{G}{k|\mathcal{P}}{}=\alpha_{k}\gss{I}{L_{h}}{},
\end{equation}
with
\[
\alpha_{k}=\sum_{l=k}^{L_{g}-1}g(l-k)g(l), \mbox{\ \ for\ } k=0,1,\ldots,L_{h}-1
\]
being the lag-$k$ autocorrelation of $g$ (decreasing with $k$) and $\alpha_{0}=1$. Moreover, the noise covariance for the active subcarriers is then equal to 
\[
\gss{C}{\sbm{\eta}|\mathcal{P}}{}=\sigma^{2}\gss{I}{L_{h}}{}
\]
Then
\[
\gss{\mathcal{G}}{\mathcal{P}}{}=\bmx{ccccc} \gss{I}{L_{h}}{} & \alpha_{1}\gss{W}{\mathcal{P}}{} & \alpha_{2}\gss{W}{\mathcal{P}}{2} & \cdots & \alpha_{L_{h}-1}\gss{W}{\mathcal{P}}{L_{h}-1} \emx
\]
and the conditions~\pref{eq:ortho1}, \pref{eq:ortho2} become
\[
\gss{d}{\mathcal{P}}{H}\gss{W}{\mathcal{P}}{k}\gss{d}{\mathcal{P}}{}=0, \mbox{\ \ } k=1,2,\ldots,L_{h}-1
\]
or equivalently
\[
\sum_{i=1}^{L_{h}}|d_{p_{i}}|^{2}e^{-j\frac{2\pi}{M}p_{i}k}=0, \mbox{\ \ } k=1,2,\ldots,L_{h}-1
\]
If the pilot symbols are taken as equipowered with $|d_{p_{i}}|^{2}=P$ for all $i$, the above is equivalent to 
\[
\sum_{i=1}^{L_{h}}e^{-j\frac{2\pi}{M}p_{i}k}=0, \mbox{\ \ } k=1,2,\ldots,L_{h}-1
\]
The latter is satisfied if additionally pilot tones $p_{i}$ are chosen to be equispaced, say $p_{i}=p_{0}+(i-1)\frac{M}{L_{h}}$ (assuming without loss of generality that $L_{h}$ divides $M$). 
Note that the constraint~\pref{eq:aBa} is now written as
\[
\|\g{d}\|^{2}\leq \mathcal{E}
\]
and is clearly satisfied with equality at an optimal point. It is clear that such an optimal solution is given by the above choice of equipowered and equispaced pilot tones, exactly as in~\cite{kkrt10}, with the $k$th diagonal entry of $\gss{\Gamma}{\mathcal{P}}{H}\gss{\Gamma}{\mathcal{P}}{}$ equal to $\alpha_{k}^{2}\|\g{d}\|^{2}$ and $|d_{p_{i}}|=\sqrt{\frac{\mathcal{E}}{L_{h}}}$. The cost function~\pref{eq:MMSE} is then given by
\begin{equation}
\mathrm{MSE}=\sigma^{2}\sum_{k=0}^{L_{h}-1}\frac{1}{\alpha_{k}^{2}\mathcal{E}}=\frac{\frac{1}{L_{h}}\left(1+\sum_{k=1}^{L_{h}-1}\frac{1}{\alpha_{k}^{2}}\right)}{\mathrm{SNR}_{\mathrm{sbc}}}
\label{eq:sparseMMSE}
\end{equation}
with $\mathrm{SNR}_{\mathrm{sbc}}=\frac{\mathcal{E}/L_{h}}{\sigma^{2}}$ denoting the per-subcarrier SNR. In view of~\pref{eq:MSEf=M*MSEt}, the MSE for CFR estimation equals $\frac{\frac{M}{L_{h}}\left(1+\sum_{k=1}^{L_{h}-1}\frac{1}{\alpha_{k}^{2}}\right)}{\mathrm{SNR}_{\mathrm{sbc}}}$. 

\medskip

\noindent
\emph{Remarks.}
\begin{enumerate}
\item Eq.~\pref{eq:hest} is now written as
\[
\g{\hat{h}}=\left(\gss{\Gamma}{\mathcal{P}}{H}\gss{\Gamma}{\mathcal{P}}{}\right)^{-1}\gss{\Gamma}{\mathcal{P}}{H}\gss{y}{\mathcal{P}}{}=\gss{\Gamma}{\mathcal{P}}{-1}\gss{y}{\mathcal{P}}{}
\]
Obviously, the $L_{h}\times L_{h}$ matrix $\gss{\Gamma}{\mathcal{P}}{-1}$ can be computed offline. The above computation then requires $L_{h}^{2}$ complex multiplications.
\item The optimal $\gss{d}{\mathcal{P}}{}$ can be written as
\[
\gss{d}{\mathcal{P}}{}=\sqrt{\frac{\mathcal{E}}{L_{h}}}\bmx{c} e^{j\theta_{0}} \\ e^{j\theta_{1}} \\ \vdots \\ e^{j\theta_{L_{h}-1}} \emx
\]
with the angles $\theta_{k}$, $k=0,1,\ldots,L_{h}-1$ being randomly chosen. For the sake of the simplicity and without loss of generality, let $p_{1}=0$. Then $p_{i}=(i-1)\frac{M}{L_{h}}$, $i=1,2,\ldots,L_{h}$ and 
\[
\gss{W}{\mathcal{P}}{}=\diag\left(1,e^{-j\frac{2\pi}{L_{h}}},e^{-j2\frac{2\pi}{L_{h}}},\ldots,e^{-j(L_{h}-1)\frac{2\pi}{L_{h}}}\right)
\]
One can then readily verify that 
\begin{equation}
\gss{\Gamma}{\mathcal{P}}{}\g{h}=\diag(\gss{d}{\mathcal{P}}{})\gss{H}{\sbm{\alpha}}{}
\label{eq:Ah}
\end{equation}
where $\gss{H}{\sbm{\alpha}}{}=\bmx{cccc} H_{\sbm{\alpha}}(0) & H_{\sbm{\alpha}}(1) & \cdots & H_{\sbm{\alpha}}(L_{h}-1) \emx^{T}$ is the $L_{h}$-point DFT of the impulse response weighted (or windowed) by the $\alpha_{k}$'s:
\[
H_{\sbm{\alpha}}(l) = \sum_{k=0}^{L_{h}-1}\alpha_{k}h(k)e^{-jkl\frac{2\pi}{L_{h}}}, \mbox{\ \ } l=0,1,\ldots,L_{h}-1
\]
with $\g{\alpha}=\bmx{cccc} \alpha_{0} & \alpha_{1} & \cdots & \alpha_{L_{h}-1} \emx^{T}$. 
Recall that $\alpha_{0}=1$ and $\alpha_{k}$ is decreasing with increasing $k$, meaning that the taps of the channel that correspond to longer delays contribute less and less. 
Moreover, note that $H_{\sbm{\alpha}}(l)$ is also the $M$-point DFT at the $l$th pilot tone. 
Eq.~\pref{eq:Ah} suggests the following computations for estimating the channel:
\begin{enumerate}
\item Compute the ``windowed'' CFR vector first:
\begin{equation}
\gss{\hat{H}}{\sbm{\alpha}}{}=\gss{y}{\mathcal{P}}{}\oslash\gss{d}{\mathcal{P}}{},
\label{eq:y/a}
\end{equation}
with $\oslash$ denoting entrywise division.
\item Compute the ``windowed'' impulse response via IFFT and divide by the weights $\alpha_{k}$'s to arrive at the CIR estimate:
\begin{equation}
\g{\hat{h}}=\mathrm{IDFT}(\gss{\hat{H}}{\sbm{\alpha}}{})\oslash \g{\alpha}
\label{eq:H/a}
\end{equation}
\end{enumerate}
\end{enumerate}
This alternative procedure costs $\frac{L_{h}}{2}\log_{2}L_{h}$ multiplications for the IDFT, $L_{h}$ complex divisions (nevertheless amenable to simplification) for~\pref{eq:y/a}, and $2(L_{h}-1)$ real divisions for~\pref{eq:H/a}. 

\begin{figure}
\begin{center}
\begin{tabular}{c}
\includegraphics[width=12cm]{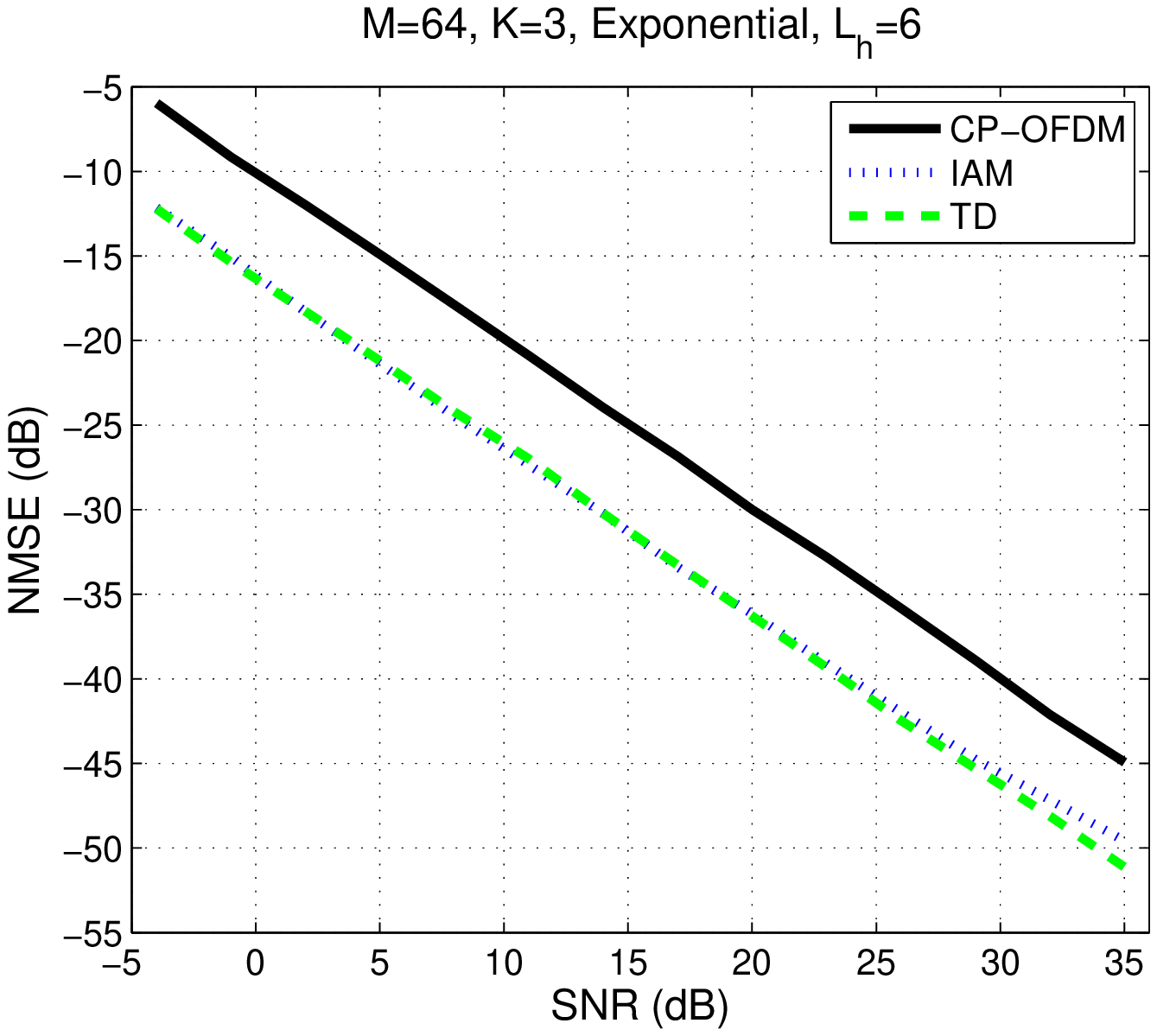} \\
(a) \\
 \includegraphics[width=12cm]{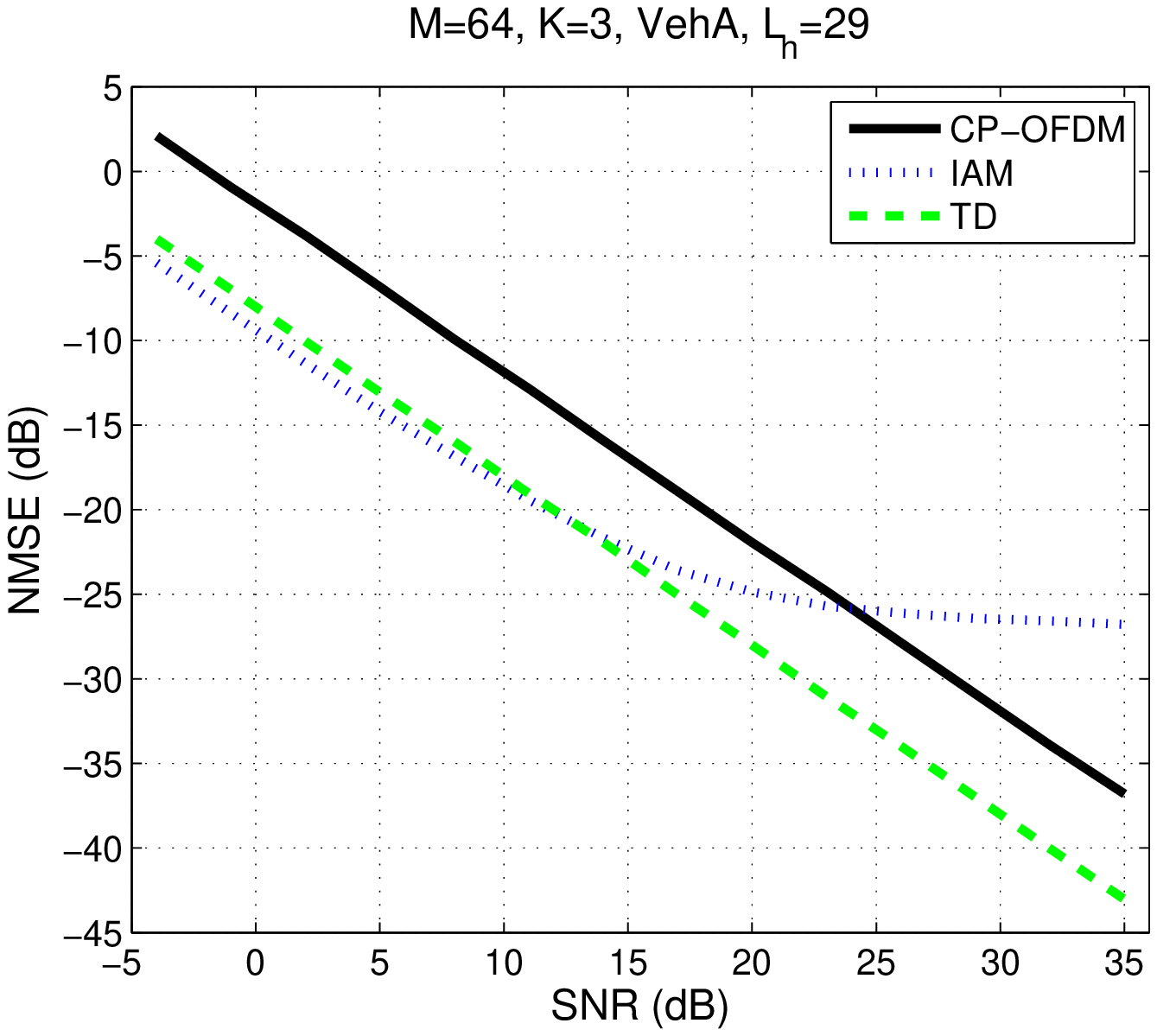} \\
 (b)
 \end{tabular}
 \end{center}
  \caption{MSE performance of the methods under study, with optimal sparse preambles, for channels of (a) low and (b) high frequency selectivity.}\label{fig:MSEsparse}
\end{figure}

Two illustrative examples of the MSE performance of the above method are given in Fig.~\ref{fig:MSEsparse}, for channels of low and high frequency selectivity.
Again, normalized MSE for CFRs is shown. 
IAM employed equispaced and equipowered pilots. The knowledge of the channel length was made use of via a DFT interpolation of the estimated CFR values.
For CP-OFDM, equispaced and equal pilots were used, in accordance with the results of~\cite{kkrt10}. Again, appropriate scaling was applied to ensure the transmit powers are the same for all methods. 
Observe how the TD method developed here is freed of the error floor effect present in the IAM method. The latter is more severe for more frequency selective channels (where the flat subchannels assumption is far from being met). Notice, however, that, for low to medium SNR values and in Fig.~\ref{fig:MSEsparse}(b), TD is slightly outperformed by IAM. 

\section{Longer Preambles}
\label{sec:longer}

So far, the preambles considered consisted of a single pilot FBMC symbol surrounded by zero guards. For highly frequency selective channels, it would be of interest to also study the case of longer preambles of this kind. For the sake of simplicity, the case of \emph{two consecutive} FBMC pilot symbols will be considered in the following. As before, a zero guard symbol is transmitted before and after this pair of pilot symbols. Let the pilots be transmitted at the time instants $n=1,2$ and call the corresponding $M$-vectors of pilots $\gss{d}{1}{}$ and $\gss{d}{2}{}$, respectively.
Then eq.~\pref{eq:ypq} is written as
\begin{eqnarray*}
y_{p,q} & = & \sum_{k=0}^{L_{h}-1}h(k)\sum_{n=1}^{2}\sum_{m=0}^{M-1}e^{-j\frac{2\pi}{M}mk}d_{m,n}j^{m+n-p-q}(-1)^{mn-pq}\sum_{l}g\left(l-k-n\frac{M}{2}\right)g\left(l-q\frac{M}{2}\right)\times \nonumber \\
& & e^{j\frac{2\pi}{M}(m-p)\left(l-\frac{L_{g}-1}{2}\right)} + \eta_{p,q},
\end{eqnarray*}
for $q=1,2$, or
\begin{equation}
y_{p,q}=\sum_{k=0}^{L_{h}-1}h(k)\sum_{n=1}^{2}\Gamma_{p,k}^{(q,n)}+\eta_{p,q},
\label{eq:ypqn}
\end{equation}
where
\begin{equation}
\Gamma_{p,k}^{(q,n)}=\sum_{m=0}^{M-1}e^{-j\frac{2\pi}{M}mk}d_{m,n}j^{m+n-p-q}(-1)^{mn-pq}\sum_{l}g\left(l-k-n\frac{M}{2}\right)g\left(l-q\frac{M}{2}\right)e^{j\frac{2\pi}{M}(m-p)\left(l-\frac{L_{g}-1}{2}\right)}
\label{eq:Gammapkqn}
\end{equation}
Clearly, 
\[
\Gamma_{p,k}^{(q,q)}=\Gamma_{p,k}(\gss{d}{q}{})=\gss{\mathcal{G}}{p,k}{H}\gss{d}{q}{},
\]
where $\Gamma_{p,k}(\gss{d}{q}{})$ denotes the $\Gamma$ quantities defined previously but with input $\gss{d}{q}{}$. 
With $q\neq n$, and recalling that there is also interference coming from adjacent times and farther than adjacent subcarriers~\cite{kkrt13}, one can write
\begin{eqnarray}
\Gamma_{p,k}^{(1,2)} & = & \sum_{m=p-2}^{p+2}d_{m,2}j^{m-p+1}(-1)^{m}e^{-j\frac{2\pi}{M}mk}e^{-j\frac{2\pi}{M}(m-p)\left(\frac{L_{g}-1}{2}\right)}\sum_{l=k+\frac{M}{2}}^{L_{g}-1}g\left(l-k-\frac{M}{2}\right)g(l)e^{j\frac{2\pi}{M}(m-p)l} \label{eq:(1,2)}\\
\Gamma_{p,k}^{(2,1)} & = & \sum_{m=p-2}^{p+2}d_{m,1}j^{m-p-1}(-1)^{p}e^{-j\frac{2\pi}{M}mk}e^{-j\frac{2\pi}{M}(m-p)\left(\frac{L_{g}-1}{2}\right)}\sum_{l}g(l-k)g\left(l-\frac{M}{2}\right)e^{j\frac{2\pi}{M}(m-p)l}
\label{eq:(2,1)}
\end{eqnarray}
It is readily seen that
\begin{eqnarray}
\Gamma_{p,k}^{(1,2)} & = & j\Gamma_{p,k+\frac{M}{2}}(\gss{d}{2}{}) \\
\Gamma_{p,k}^{(2,1)} & = & -j\Gamma_{p,k-\frac{M}{2}}(\gss{d}{1}{})
\end{eqnarray}
More generally,
\begin{equation}
\Gamma_{p,k}^{(q,n)} = j^{n-q}\Gamma_{p,k+(n-q)\frac{M}{2}}(\gss{d}{n}{})
\end{equation}

In view of the above, \pref{eq:ypqn} can be written more compactly as
\begin{equation}
\gss{y}{q}{}=(\gss{\Gamma}{}{(q,1)}+\gss{\Gamma}{}{(q,2)})\g{h}+\gss{\eta}{q}{}, \mbox{\ \ \ } q=1,2
\label{eq:yp12}
\end{equation}
with the obvious definition for matrices $\gss{\Gamma}{}{(q,n)}$ and with $\gss{y}{q}{}=\bmx{cccc} y_{0,q} & y_{1,q} & \cdots & y_{M-1,q} \emx^{T}$ and $\gss{\eta}{q}{}$ defined analogously. 
Moreover, one can write expressions analogous to~\pref{eq:A=GD}, that is,
\begin{equation}
\gss{\Gamma}{}{(q,n)}=\gss{\mathcal{G}}{}{(q,n)}\gss{D}{n}{}
\end{equation}
with 
\[
\gss{D}{n}{}=\gss{I}{L_{h}}{}\otimes \gss{d}{n}{}
\]
and obvious definition for $\gss{\mathcal{G}}{}{(q,n)}$. 
Hence, with the equations~\pref{eq:yp12} put together, the following results
\[
\underbrace{\bmx{c} \gss{y}{1}{} \\ \gss{y}{2}{} \emx}_{\g{\bar{y}}} = \underbrace{\bmx{cccc} \g{\mathcal{G}} & \gss{\mathcal{G}}{}{(1,2)} \\ \gss{\mathcal{G}}{}{(2,1)} & \g{\mathcal{G}} \emx}_{\g{\bar{\mathcal{G}}}} \underbrace{\bmx{c} \gss{D}{1}{} \\ \gss{D}{2}{} \emx}_{\g{\bar{D}}}\g{h}+\underbrace{\bmx{c} \gss{\eta}{1}{} \\ \gss{\eta}{2}{}\emx}_{\g{\bar{\eta}}}
\]
or 
\begin{equation}
\g{\bar{y}}=\g{\bar{\Gamma}}\g{h}+\g{\bar{\eta}},
\end{equation}
where
\begin{equation}
\g{\bar{\Gamma}}=\g{\bar{\mathcal{G}}}\g{\bar{D}}
\end{equation}
The estimate for $\g{h}$ then results similarly with~\pref{eq:hest}:
\begin{equation}
\g{\hat{h}}=(\gss{\bar{\Gamma}}{}{H}\gss{C}{\sbm{\bar{\eta}}}{-1}\g{\bar{\Gamma}})^{-1}\gss{\bar{\Gamma}}{}{H}\gss{C}{\sbm{\bar{\eta}}}{-1}\g{\bar{y}},
\label{eq:hbarest}
\end{equation}
where $\gss{C}{\sbm{\bar{\eta}}}{}$ is the covariance matrix of $\g{\bar{\eta}}$. It can be readily shown to be given by
\begin{equation}
\gss{C}{\sbm{\bar{\eta}}}{}=\sigma^{2}\bmx{cc} \g{B} & \g{S}\gss{A}{}{+} \\ \g{S}\gss{A}{}{-} & \g{B} \emx \equiv \sigma^{2}\g{\bar{B}}
\label{eq:Chbar}
\end{equation}
where 
\[
\g{S}=\diag(1,-1,1,-1,\ldots,1,-1)
\]
and the matrices $\gss{A}{}{\pm}$ are banded and symmetric circulant\footnote{Except for some minor discrepancies at the boundaries of the matrix, which can be overlooked for large enough $M$.}:
\[
\gss{A}{}{\pm}=j
\bmx{cccccccccc} 
\pm\gamma & \delta & \mp\epsilon & 0 & 0 & 0 & 0 & \cdots & \pm\epsilon & -\delta \\ 
\delta & \pm\gamma & \delta & \mp\epsilon & 0 & 0 & 0 & \cdots & 0 & \mp\epsilon \\
\mp\epsilon & \delta & \pm\gamma & \delta & \mp\epsilon & 0 & 0 & \cdots & 0 & 0 \\
0 & \mp\epsilon & \delta & \pm\gamma & \delta & \mp\epsilon & 0 & \cdots & 0 & 0 \\
\vdots & \ddots & \ddots & \ddots & \ddots & \ddots & \ddots & \vdots & \ddots & \ddots \\
-\delta & \pm\epsilon & 0 & 0 & 0 & 0 & 0 & \cdots & \delta & \pm\gamma 
\emx
\]
The quantities $\gamma,\delta,\epsilon$ are \emph{a priori} computable from the knowledge of the prototype filter $g$ and their expressions are given in~\cite{kkrt13}. By the way, recall that (in general) $\gamma > \delta \gg |\epsilon|$. 
In the circulant matrix terminology, the matrices $\g{S}\gss{A}{}{\pm}$ above are known as \emph{alternating circulant}~\cite{t05}. Thus, the matrix in~\pref{eq:Chbar} is Hermitian\footnote{But not block circulant. This makes the following derivation of its eigen-decomposition far from being trivial.} with circulant diagonal and alternating circulant off-diagonal blocks. 
Furthermore, one can show that, in line with~\pref{eq:Gcalk}, the $\gss{\mathcal{G}}{}{(q,n)}$ matrices above can be written as
\begin{equation}
\gss{\mathcal{G}}{}{(q,n)} = \g{S}\bmx{ccccc} \gss{G}{0}{(q,n)} & \g{W}\gss{G}{1}{(q,n)} & \gss{W}{}{2}\gss{G}{2}{(q,n)} \cdots & \gss{W}{}{L_{h}-1}\gss{G}{L_{h}-1}{(q,n)} \emx,
\end{equation}
where $\g{W}$ is defined as previously and the matrices $\gss{G}{k}{(q,n)}$ are banded and circulant. In fact, they are (approximately) tridiagonal and symmetric, with imaginary entries.
Specifically, as it can be seen from eqs.~\pref{eq:(1,2)} and~\pref{eq:(2,1)}, 
\begin{eqnarray}
\gss{G}{0}{(1,2)} & = & \gss{A}{}{+} \\
\gss{G}{0}{(2,1)} & = & \gss{A}{}{-}
\end{eqnarray}
For example, 
\begin{eqnarray*}
\Gamma_{p,0}^{(1,2)} & = & d_{p-2,2}j^{-1}(-1)^{p}e^{j2\frac{2\pi}{M}\left(\frac{L_{g}-1}{2}\right)}\sum_{l}g(l)g\left(l-\frac{M}{2}\right)e^{-j2\frac{2\pi}{M}l} + \\
& & d_{p-1,2}(-1)^{p-1}e^{j\frac{2\pi}{M}\left(\frac{L_{g}-1}{2}\right)}\sum_{l}g(l)g\left(l-\frac{M}{2}\right)e^{-j\frac{2\pi}{M}l} + \\
& & d_{p,2}j(-1)^{p}\sum_{l}g(l)g\left(l-\frac{M}{2}\right)+ \\
& & d_{p+1,2}(-1)^{p}e^{-j\frac{2\pi}{M}\left(\frac{L_{g}-1}{2}\right)}\sum_{l}g(l)g\left(l-\frac{M}{2}\right)e^{j\frac{2\pi}{M}l}+ \\
& & d_{p+2,2}j^{3}(-1)^{p}e^{-j2\frac{2\pi}{M}\left(\frac{L_{g}-1}{2}\right)}\sum_{l}g(l)g\left(l-\frac{M}{2}\right)e^{j2\frac{2\pi}{M}l} \\
& = & -j(-1)^{p}\epsilon d_{p-2,2}+j(-1)^{p}\delta d_{p-1,2}+j(-1)^{p}\gamma d_{p,2}+j(-1)^{p}\delta d_{p+1,2}-j(-1)^{p}\epsilon d_{p+2,2}
\end{eqnarray*}

The aim now is to diagonalize the above noise covariance matrix so that the Gauss-Markov estimator in~\pref{eq:hbarest} can be analyzed and its optimal preamble properly designed.
For that purpose, let 
\[
\gss{A}{}{\pm}=\g{F}\gss{\Lambda}{}{\pm}\gss{F}{}{H}
\]
be the eigenvalue decompositions of the (circulant and symmetric) matrices $\gss{A}{}{\pm}$. Note that their eigenvalues are imaginary valued. 
Note also that 
\[
\g{S}=\gss{W}{}{M/2}
\]
Hence, invoking~\pref{eq:FW=ZF} with $k=\frac{M}{2}$ and using the eigenvalue decompositions above, the following results\footnote{This can also be deduced from the circular convolution property of the DFT.}
\[
\g{S}\gss{A}{}{\pm}=\g{F}\gss{Z}{}{M/2}\gss{\Lambda}{}{\pm}\gss{F}{}{H},
\]
where
\[
\gss{Z}{}{M/2}\gss{\Lambda}{}{\pm}=\bmx{cc} \gss{0}{\frac{M}{2}\times \frac{M}{2}}{} & \gssa{\Lambda}{}{\pm}{M/2+1:M,M/2+1:M} \\ \gssa{\Lambda}{}{\pm}{1:M/2,1:M/2} & \gss{0}{\frac{M}{2}\times \frac{M}{2}}{} \emx \equiv \gss{\Lambda}{Z}{\pm}
\]
Hence, $\g{\bar{B}}$ in~\pref{eq:Chbar} can be expressed as
\begin{eqnarray}
\g{\bar{B}} & = & \bmx{cc} \g{F}\g{\Lambda}\gss{F}{}{H} & \g{F}\gss{\Lambda}{Z}{+}\gss{F}{}{H} \\ \g{F}\gss{\Lambda}{Z}{-}\gss{F}{}{H} & \g{F}\g{\Lambda}\gss{F}{}{H} \emx \nonumber \\
& = & (\gss{I}{2}{}\otimes \g{F})\underbrace{\bmx{cc} \g{\Lambda} & \gss{\Lambda}{Z}{+} \\ \gss{\Lambda}{Z}{-} & \g{\Lambda} \emx}_{\g{K}} (\gss{I}{2}{}\otimes \g{F})^{H}
\label{eq:Bbar}
\end{eqnarray}
Since $\gss{\Lambda}{}{\pm}$ are diagonal with imaginary entries, there holds
\[
\gss{\Lambda}{Z}{-}=(\gss{\Lambda}{Z}{+})^{H}=-\gss{\Lambda}{}{+}\gss{Z}{}{M/2}
\]
and 
\begin{equation}
\gss{\Lambda}{}{-}=-\gss{Z}{}{M/2}\gss{\Lambda}{}{+}\gss{Z}{}{M/2}
\label{eq:L-}
\end{equation}
Hence the matrix $\g{K}$ in~\pref{eq:Bbar} can be written as
\begin{eqnarray}
\g{K} & = & \bmx{cc} \g{\Lambda} & \gss{Z}{}{M/2}\gss{\Lambda}{}{+} \\ -\gss{\Lambda}{}{+}\gss{Z}{}{M/2}\emx \nonumber \\
& = & \bmx{cc} \gss{Z}{}{M/2} & \g{0} \\ \g{0} & -j\gss{I}{M}{} \emx
\bmx{cc} \gss{\Lambda}{}{\prime} & \gss{\Lambda}{I}{+} \\ \gss{\Lambda}{I}{+} & \g{\Lambda} \emx \bmx{cc} \gss{Z}{}{M/2} & \g{0} \\ \g{0} & -j\gss{I}{M}{} \emx^{H}, \label{eq:K1}
\end{eqnarray}
with
\[
\gss{\Lambda}{}{\prime}=\gss{Z}{}{M/2}\g{\Lambda}\gss{Z}{}{M/2} 
\]
and
\[
\gss{\Lambda}{I}{+}=\Im\{\gss{\Lambda}{}{+}\}
\]
Using in~\pref{eq:K1} the symmetries of the matrices $\g{\Lambda}$ and $\gss{\Lambda}{I}{+}$, namely
\begin{eqnarray*}
\g{\Lambda} & = & \bmx{cc} \gss{J}{M/2+1}{} & \g{0} \\ \g{0} & \gss{J}{M/2-1}{} \emx \g{\Lambda} \bmx{cc} \gss{J}{M/2+1}{} & \g{0} \\ \g{0} & \gss{J}{M/2-1}{} \emx, \\
\gss{\Lambda}{I}{+} & = & \underbrace{\bmx{cc} 1 & \g{0} \\ \g{0} & \gss{J}{M-1}{} \emx}_{\g{P}} \gss{\Lambda}{I}{+} \underbrace{\bmx{cc} 1 & \g{0} \\ \g{0} & \gss{J}{M-1}{} \emx}_{\g{P}},
\end{eqnarray*}
with $\gss{J}{N}{}$ being the $N$th order antidiagonal (exchange) matrix, the following results
\begin{equation}
\g{K}=
\bmx{cc} \gss{Z}{}{M/2} & \g{0} \\ \g{0} & -j\gss{I}{M}{} \emx
\bmx{cc} \g{P} & \g{0} \\ \g{0} & \gss{I}{M}{} \emx \underbrace{\bmx{cc} \g{\Lambda} & \g{P}\gss{\Lambda}{I}{+} \\ \g{P}\gss{\Lambda}{I}{+} & \g{\Lambda} \emx}_{\g{M}}
\bmx{cc} \g{P} & \g{0} \\ \g{0} & \gss{I}{M}{} \emx
\bmx{cc} \gss{Z}{}{M/2} & \g{0} \\ \g{0} & -j\gss{I}{M}{} \emx^{H} \label{eq:K2}
\end{equation}
The matrix $\g{M}$ is \emph{block circulant}~\cite{t05} and hence can be block-diagonalized with the aid of the matrix $\gss{F}{2}{}\otimes \gss{I}{M}{}$, where $\gss{F}{2}{}=\frac{1}{\sqrt{2}}\bmx{cc} 1 & 1 \\ 1 & -1 \emx$ is the $2\times 2$ unitary DFT matrix:
\begin{equation}
\g{M}=\frac{1}{2}\bmx{cr} \g{I} & \g{I} \\ \g{I} & -\g{I} \emx\,\underbrace{\bmx{cc} \g{\Lambda}+\g{P}\gss{\Lambda}{I}{+} & \g{0} \\ \g{0} & \g{\Lambda}-\g{P}\gss{\Lambda}{I}{+} \emx}_{\g{N}}\,\bmx{cr} \g{I} & \g{I} \\ \g{I} & -\g{I} \emx
\label{eq:M}
\end{equation}
Each of the diagonal blocks $\g{\Lambda}\pm\g{P}\gss{\Lambda}{I}{+}$ can be expressed as the product of $\frac{M}{2}+1$ symmetric matrices, which are simple modifications of the identity matrix. 
For example, with $M=8$ and $\lambda_{i}^{+}$ denoting the diagonal entries of $\gss{\Lambda}{I}{+}$:
\begin{eqnarray*}
\gss{N}{}{\pm} \equiv \g{\Lambda}\pm\g{P}\gss{\Lambda}{I}{+} & = & 
\bmx{cccccccc} \lambda_{1}\pm\lambda_{1}^{+} & 0 & 0 & 0 & 0 & 0 & 0 & 0 \\
0 & \lambda_{2} & 0 & 0 & 0 & 0 & 0 & \pm\lambda_{2}^{+} \\
0 & 0 & \lambda_{3} & 0 & 0 & 0 & \pm\lambda_{3}^{+} & 0 \\
0 & 0 & 0 & \lambda_{2} & 0 & \pm\lambda_{4}^{+} & 0 & 0 \\
0 & 0 & 0 & 0 & \lambda_{1}\pm\lambda_{5}^{+} & 0 & 0 & 0 \\
0 & 0 & 0 & \pm\lambda_{4}^{+} & 0 & \lambda_{6} & 0 & 0 \\
0 & 0 & \pm\lambda_{3}^{+} & 0 & 0 & 0 & \lambda_{7} & 0 \\
0 & \pm\lambda_{2}^{+} & 0 & 0 & 0 & 0 & 0 & \lambda_{6} 
\emx \\
& = & 
\bmx{cccccccc} \lambda_{1}\pm\lambda_{1}^{+} & 0 & 0 & 0 & 0 & 0 & 0 & 0 \\
0 & 1 & 0 & 0 & 0 & 0 & 0 & 0 \\
0 & 0 & 1 & 0 & 0 & 0 & 0 & 0 \\
0 & 0 & 0 & 1 & 0 & 0 & 0 & 0 \\
0 & 0 & 0 & 0 & 1 & 0 & 0 & 0 \\
0 & 0 & 0 & 0 & 0 & 1 & 0 & 0 \\
0 & 0 & 0 & 0 & 0 & 0 & 1 & 0 \\
0 & 0 & 0 & 0 & 0 & 0 & 0 & 1 
\emx
\bmx{cccccccc} 1 & 0 & 0 & 0 & 0 & 0 & 0 & 0 \\
0 & \lambda_{2} & 0 & 0 & 0 & 0 & 0 & \pm\lambda_{2}^{+} \\
0 & 0 & 1 & 0 & 0 & 0 & 0 & 0 \\
0 & 0 & 0 & 1 & 0 & 0 & 0 & 0 \\
0 & 0 & 0 & 0 & 1 & 0 & 0 & 0 \\
0 & 0 & 0 & 0 & 0 & 1 & 0 & 0 \\
0 & 0 & 0 & 0 & 0 & 0 & 1 & 0 \\
0 & \pm\lambda_{2}^{+} & 0 & 0 & 0 & 0 & 0 & \lambda_{6} 
\emx \times \\
& & \bmx{cccccccc} 1 & 0 & 0 & 0 & 0 & 0 & 0 & 0 \\
0 & 1 & 0 & 0 & 0 & 0 & 0 & 0 \\
0 & 0 & \lambda_{3} & 0 & 0 & 0 & \pm\lambda_{3}^{+} & 0 \\
0 & 0 & 0 & 1 & 0 & 0 & 0 & 0 \\
0 & 0 & 0 & 0 & 1 & 0 & 0 & 0 \\
0 & 0 & 0 & 0 & 0 & 1 & 0 & 0 \\
0 & 0 & \pm\lambda_{3}^{+} & 0 & 0 & 0 & \lambda_{7} & 0 \\
0 & 0 & 0 & 0 & 0 & 0 & 0 & 1 
\emx
\bmx{cccccccc} 1 & 0 & 0 & 0 & 0 & 0 & 0 & 0 \\
0 & 1 & 0 & 0 & 0 & 0 & 0 & 0 \\
0 & 0 & 1 & 0 & 0 & 0 & 0 & 0 \\
0 & 0 & 0 & \lambda_{2} & 0 & \pm\lambda_{4}^{+} & 0 & 0 \\
0 & 0 & 0 & 0 & 1 & 0 & 0 & 0 \\
0 & 0 & 0 & \pm\lambda_{4}^{+} & 0 & \lambda_{6} & 0 & 0 \\
0 & 0 & 0 & 0 & 0 & 0 & 1 & 0 \\
0 & 0 & 0 & 0 & 0 & 0 & 0 & 1 
\emx \times \\
& & \bmx{cccccccc} 1 & 0 & 0 & 0 & 0 & 0 & 0 & 0 \\
0 & 1 & 0 & 0 & 0 & 0 & 0 & 0 \\
0 & 0 & 1 & 0 & 0 & 0 & 0 & 0 \\
0 & 0 & 0 & 1 & 0 & 0 & 0 & 0 \\
0 & 0 & 0 & 0 & \lambda_{1}\pm\lambda_{5}^{+} & 0 & 0 & 0 \\
0 & 0 & 0 & 0 & 0 & 1 & 0 & 0 \\
0 & 0 & 0 & 0 & 0 & 0 & 1 & 0 \\
0 & 0 & 0 & 0 & 0 & 0 & 0 & 1 
\emx \\
& \equiv & \gss{N}{1}{\pm}\gss{N}{2}{\pm}\gss{N}{3}{\pm}\gss{N}{4}{\pm}\gss{N}{5}{\pm},
\end{eqnarray*}
where the symmetries of the eigenvalues outlined above have also been taken into account. 
Moreover, note that these factors commute and can thus be multiplied in any order. Each of them is basically a $2\times 2$ real symmetric matrix, embedded into the $M$th-order identity. Hence, they can be eigen-decomposed via Givens rotations. Obviously, in addition to their unity eigenvalues, these matrices have eigenvalues as follows:
\begin{eqnarray*}
\gss{N}{1}{\pm}: & & \lambda_{1}\pm\lambda_{1}^{+}, \\
\gss{N}{2}{\cdot}: & & \frac{\lambda_{2}+\lambda_{6}\pm\sqrt{(\lambda_{2}-\lambda_{6})^{2}+(2\lambda_{2}^{+})^{2}}}{2}, \\
\gss{N}{3}{\cdot}: & & \frac{\lambda_{3}+\lambda_{7}\pm\sqrt{(\lambda_{3}-\lambda_{7})^{2}+(2\lambda_{3}^{+})^{2}}}{2}, \\
\gss{N}{4}{\cdot}: & & \frac{\lambda_{2}+\lambda_{6}\pm\sqrt{(\lambda_{2}-\lambda_{6})^{2}+(2\lambda_{4}^{+})^{2}}}{2}, \\
\gss{N}{5}{\pm}: & & \lambda_{1}\pm\lambda_{5}^{+},
\end{eqnarray*}
with corresponding eigen-matrices
\begin{eqnarray*}
\gss{V}{1}{\pm} & = & \gss{I}{8}{}, \\
\gss{V}{2}{\pm} & = & 
\bmx{cccccccc} 1 & 0 & 0 & 0 & 0 & 0 & 0 & 0 \\ 
0 & c_{2} & 0 & 0 & 0 & 0 & 0 & \pm s_{2} \\
0 & 0 & 1 & 0 & 0 & 0 & 0 & 0 \\
0 & 0 & 0 & 1 & 0 & 0 & 0 & 0 \\
0 & 0 & 0 & 0 & 1 & 0 & 0 & 0 \\
0 & 0 & 0 & 0 & 0 & 1 & 0 & 0 \\
0 & 0 & 0 & 0 & 0 & 0 & 1 & 0 \\
0 & \mp s_{2} & 0 & 0 & 0 & 0 & 0 & c_{2}
\emx=\ga{G}{2,8,\pm\theta_{2}}, \\
\gss{V}{3}{\pm} & = & \ga{G}{3,7,\pm\theta_{3}}, \\
\gss{V}{4}{\pm} & = & \ga{G}{4,6,\pm\theta_{4}}, \\
\gss{V}{5}{\pm} & = & \gss{I}{8}{},
\end{eqnarray*}
where the notation $\ga{G}{i,k,\theta}$ has been used to denote the $M\times M$ Givens rotation matrix with $c=\cos(\theta)$ at its $(i,i)$ and $(k,k)$ entries, and $s=\sin(\theta)$ and $-s$ at its $(i,k)$ and $(k,i)$ entries, respectively~\cite{gv89}. The angles $\theta_{i}$ can be computed on the basis of the entries of $\gss{N}{i}{\cdot}$'s. Notice that the $\gss{V}{i}{}$ matrices are real-valued and 
\[
\gss{V}{i}{+}=(\gss{V}{i}{-})^{-1}=(\gss{V}{i}{-})^{T}
\]
Moreover, similarly with the $\gss{N}{i}{}$ matrices, the above eigen-matrices can be seen to commute. It is then straightforward to verify that the $\gss{N}{\pm}{}$ matrix has eigenvalues 
as above, placed on the main diagonal of the diagonal matrix $\gss{L}{}{\pm}$, and eigen-matrix
\[
\gss{V}{}{\pm}=\prod_{i=1}^{M/2+1}\gss{V}{i}{\pm}=\prod_{i=2}^{M/2}\gss{V}{i}{\pm},
\]
with $\gss{V}{}{-}=(\gss{V}{}{+})^{T}$. 
In view of the above, one can write
\begin{equation}
\g{N}=\bmx{cc} \gss{N}{}{+} & \g{0} \\ \g{0} & \gss{N}{}{-} \emx = \bmx{cc} \gss{V}{}{+} & \g{0} \\ \g{0} & (\gss{V}{}{+})^{T} \emx \bmx{cc} \gss{L}{}{+} & \g{0} \\ \g{0} & \gss{L}{}{-} \emx \bmx{cc} (\gss{V}{}{+})^{T} & \g{0} \\ \g{0} & \gss{V}{}{+}\emx
\label{eq:N}
\end{equation}
The matrices $\gss{L}{}{\pm}$ differ only in their first and $\left(\frac{M}{2}+1\right)$st diagonal entries, which equal $\lambda_{1}\pm\lambda_{1}^{+}$ and $\lambda_{1}\pm\lambda_{M/2+1}^{+}$, respectively. They agree in all other entries.  

Combining eqs.~\pref{eq:Bbar}, \pref{eq:K2}, \pref{eq:M}, and~\pref{eq:N} leads to the sought for eigen-decomposition of the noise covariance matrix in~\pref{eq:Chbar}. 
One can then write the matrix $\gss{\bar{\Gamma}}{}{H}\gss{C}{\sbm{\bar{\eta}}}{-1}\g{\bar{\Gamma}}$ in~\pref{eq:hbarest} as $\frac{1}{\sigma^{2}}\gss{\tilde{\bar{\Gamma}}}{}{H}\g{\tilde{\bar{\Gamma}}}$, where
\begin{equation}
\g{\tilde{\bar{\Gamma}}}= \label{eq:tGamma} 
\frac{1}{\sqrt{2}}\bmx{cc} (\gss{L}{}{+})^{-1/2} & \g{0} \\ \g{0} & (\gss{L}{}{-})^{-1/2} \emx \bmx{cc} (\gss{V}{}{+})^{T} & \g{0} \\ \g{0} & \gss{V}{}{+} \emx
\bmx{cr} \g{I} & \g{I} \\ \g{I} & -\g{I} \emx \bmx{cc} \g{P} & \g{0} \\ \g{0} & \g{I} \emx \bmx{cc} \gss{Z}{}{M/2} & \g{0} \\ \g{0} & j\g{I} \emx \bmx{cc} \gss{F}{}{H} & \g{0} \\ \g{0} & \gss{F}{}{H} \emx \g{\bar{\Gamma}}
\end{equation}
In line with the single-symbol case, the energy of the SFB-modulated preamble signal can be seen to equal $\gss{\bar{d}}{}{H}\g{\bar{B}}\g{\bar{d}}$, where 
\[
\g{\bar{d}}=\bmx{cc} \gss{d}{1}{T} & \gss{d}{2}{T} \emx^{T}
\]
and $\g{\bar{B}}$ is given in~\pref{eq:Chbar}. Consider then the constrained minimization problem
\begin{eqnarray}
& & \min_{\sbm{\bar{d}}}\;\mathrm{tr}\{(\gss{\tilde{\bar{\Gamma}}}{}{H}\g{\tilde{\bar{\Gamma}}})^{-1}\} \\
\mathrm{s.t.} & & \gss{\bar{d}}{}{H}\g{\bar{B}}\g{\bar{d}}\leq \mathcal{E}
\end{eqnarray}
Clearly, the training symbols $\gss{d}{1}{},\gss{d}{2}{}$ must be so chosen as to yield a matrix $\g{\tilde{\bar{\Gamma}}}$ with orthogonal columns. 
Consider the transformation in~\pref{eq:tGamma}. 
 First,
\begin{eqnarray*}
\lefteqn{\bmx{cc} \gss{F}{}{H} & \g{0} \\ \g{0} & \gss{F}{}{H} \emx \g{\bar{\mathcal{G}}}\g{\bar{D}}=} \\
& & \bmx{cc} \bmx{cccc} \g{\Lambda} & \g{Z}\gss{\Lambda}{1}{} & \cdots & \gss{Z}{}{L_{h}-1}\gss{\Lambda}{L_{h}-1}{} \emx & 
\gss{Z}{}{M/2}\bmx{cccc} \gss{\Lambda}{}{+} & \g{Z}\gss{\Lambda}{1}{(1,2)} & \cdots & \gss{Z}{}{L_{h}-1}\gss{\Lambda}{L_{h}-1}{(1,2)} \emx \\
\gss{Z}{}{M/2}\bmx{cccc} \gss{\Lambda}{}{-} & \g{Z}\gss{\Lambda}{1}{(2,1)} & \cdots & \gss{Z}{}{L_{h}-1}\gss{\Lambda}{L_{h}-1}{(2,1)} \emx & 
\bmx{cccc} \g{\Lambda} & \g{Z}\gss{\Lambda}{1}{} & \cdots & \gss{Z}{}{L_{h}-1}\gss{\Lambda}{L_{h}-1}{} \emx
\emx \times \\
& & \bmx{c} \gss{I}{L_{h}}{}\otimes \gss{\tilde{d}}{1}{} \\ \gss{I}{L_{h}}{}\otimes \gss{\tilde{d}}{2}{} \emx,
\end{eqnarray*}
where $\gss{\tilde{d}}{n}{}=\gss{F}{}{H}\gss{d}{n}{}$ as previously.
One can easily see that 
\[
\g{P}\gss{Z}{}{M/2}=\bmx{cc} \gss{J}{M/2+1}{} & \g{0} \\ \g{0} & \gss{J}{M/2-1}{} \emx = \gss{Z}{}{M/2}\g{P}
\]
and
\[
\g{P}\gss{Z}{}{M/2}\gss{Z}{}{k}=\bmx{cc} \gss{J}{M/2+1-k}{} & \g{0} \\ \g{0} & \gss{J}{M/2-1+k}{} \emx = \gss{Z}{}{M/2-k}\g{P}
\]
More generally,
\[
\g{P}\gss{Z}{}{k}=\bmx{cc} \gss{J}{M-k+1}{} & \g{0} \\ \g{0} & \gss{J}{k-1}{} \emx = \gss{Z}{}{-k}\g{P}
\]
Note that 
\[
\gss{\Lambda}{k}{} = \bmx{cc} \gss{J}{M/2+1-k}{} & \g{0} \\ \g{0} & \gss{J}{M/2-1+k}{} \emx \gss{\Lambda}{k}{} \bmx{cc} \gss{J}{M/2+1-k}{} & \g{0} \\ \g{0} & \gss{J}{M/2-1+k}{} \emx,
\]
which also holds for $\gss{\Lambda}{0}{}=\g{\Lambda}$.
As for the $\gss{\Lambda}{k}{(\cdot,\cdot)}$ matrices, they are diagonal with imaginary diagonal entries and the same symmetry with that of $\gss{\Lambda}{}{\pm}$ holds, namely
\[
\gss{\Lambda}{k}{(\cdot,\cdot)} = \g{P}\gss{\Lambda}{k}{(\cdot,\cdot)}\g{P}
\]
Using~\pref{eq:L-} and the above identities, one can see -- after some algebra -- that the first column of the matrix $\g{\tilde{\bar{\Gamma}}}$ equals
\begin{equation}
\ga{\tilde{\bar{\Gamma}}}{:,1}=\frac{1}{\sqrt{2}}\bmx{c} (\gss{L}{}{+})^{1/2}(\gss{V}{}{+})^{T}(\g{P}\gss{Z}{}{M/2}\gss{\tilde{d}}{1}{}+j\gss{\tilde{d}}{2}{}) \\
(\gss{L}{}{-})^{1/2}\gss{V}{}{+}(\g{P}\gss{Z}{}{M/2}\gss{\tilde{d}}{1}{}-j\gss{\tilde{d}}{2}{}) \emx
\label{eq:tGamma1}
\end{equation}
and its $(k+1)$st column, $k=1,2,\ldots,L_{h}-1$:
\begin{eqnarray}
\ga{\tilde{\bar{\Gamma}}}{:,k+1} & = & \frac{1}{\sqrt{2}}\bmx{c} (\gss{L}{}{+})^{-1/2}(\gss{V}{}{+})^{T}\left[(\gss{\Lambda}{k}{}\g{P}\gss{Z}{}{M/2+k}-\gss{Z}{}{M/2+k}\gss{\Lambda}{k,I}{(2,1)})\gss{\tilde{d}}{1}{}+\jmath (\gss{Z}{}{k}\gss{\Lambda}{k}{}+\g{P}\gss{Z}{}{k}\gss{\Lambda}{k,I}{(1,2)})\gss{\tilde{d}}{2}{}\right] \\
(\gss{L}{}{-})^{-1/2}\gss{V}{}{+}\left[(\gss{\Lambda}{k}{}\g{P}\gss{Z}{}{M/2+k}+\gss{Z}{}{M/2+k}\gss{\Lambda}{k,I}{(2,1)})\gss{\tilde{d}}{1}{}-\jmath (\gss{Z}{}{k}\gss{\Lambda}{k}{}-\g{P}\gss{Z}{}{k}\gss{\Lambda}{k,I}{(1,2)})\gss{\tilde{d}}{2}{}\right]
\emx \nonumber \\
& = & \frac{1}{\sqrt{2}}\bmx{c} (\gss{L}{}{+})^{-1/2}(\gss{V}{}{+})^{T}\bmx{cc} \gss{\Lambda}{k}{}\g{P}\gss{Z}{}{M/2+k}-\gss{Z}{}{M/2+k}\gss{\Lambda}{k,I}{(2,1)} & \jmath \left(\gss{Z}{}{k}\gss{\Lambda}{k}{}+\g{P}\gss{Z}{}{k}\gss{\Lambda}{k,I}{(1,2)}\right)\emx \\
(\gss{L}{}{-})^{-1/2}\gss{V}{}{+}\bmx{cc} \gss{\Lambda}{k}{}\g{P}\gss{Z}{}{M/2+k}+\gss{Z}{}{M/2+k}\gss{\Lambda}{k,I}{(2,1)} & -\jmath \left(\gss{Z}{}{k}\gss{\Lambda}{k}{}-\g{P}\gss{Z}{}{k}\gss{\Lambda}{k,I}{(1,2)}\right)\emx
\emx \g{\tilde{\bar{d}}}, \nonumber \\
& & \label{eq:tGammak}
\end{eqnarray}
where
\[
\g{\tilde{\bar{d}}}=\bmx{cc} \gss{\tilde{d}}{1}{T} & \gss{\tilde{d}}{2}{T} \emx^{T}
\]
The requirement for this matrix to have orthogonal columns leads to $L_{h}-1$ conditions of the form
\begin{equation}
\gssa{\tilde{\bar{\Gamma}}}{}{H}{:,1}\ga{\tilde{\bar{\Gamma}}}{:,k+1}=0, \;\; k=1,2,\ldots,L_{h}-1,
\label{eq:0k}
\end{equation}
and the $\frac{(L_{h}-1)(L_{h}-2)}{2}$ conditions
\begin{equation}
\gssa{\tilde{\bar{\Gamma}}}{}{H}{:,k+1}\ga{\tilde{\bar{\Gamma}}}{:,l+1}=0, \;\; l>k=1,2,\ldots,L_{h}-2
\label{eq:kl}
\end{equation}
Based on the previous analysis, one can show that~\pref{eq:0k} takes the form
\begin{equation}
\gss{\tilde{\bar{d}}}{}{H}(\gss{I}{2}{}\otimes \gss{Z}{}{k})\bmx{cc} \gss{\Lambda}{k}{} & \gss{Z}{}{M/2}\gss{\Lambda}{k}{(1,2)} \\ \gss{Z}{}{M/2}\gss{\Lambda}{k}{(2,1)} & \gss{\Lambda}{k}{} \emx \g{\tilde{\bar{d}}}=0,
\;\; k=1,2,\ldots,L_{h}-1
\label{eq:0ka}
\end{equation}

\appendix
\section{A Special Case: Flat Subchannels}
\label{sec:flat}

Known results for channel estimation with the assumption of the CFR being flat in each subchannel (see~\cite{kkrt13}) can be easily derived as a special case of the above. Indeed, in that case, one can write 
\[
\alpha_{k}=\sum_{l=k}^{L_{g}-1}g(l-k)g(l)\approx \sum_{l=0}^{L_{g}-1}g^{2}(l)=1, \mbox{\ \ for\ } k=0,1,\ldots,L_{h}-1
\]
Then it can be easily verified that all $\gss{G}{k}{}$ equal $\gss{G}{0}{}=\g{B}$ and hence 
\[
\g{\mathcal{G}}=\bmx{ccccc} \g{B} & \g{W}\g{B} & \gss{W}{}{2}\g{B} & \cdots & \gss{W}{}{L_{h}-1}\g{B} \emx,
\]
yielding
\begin{equation}
\g{\Gamma}=\bmx{ccccc} \g{c} & \g{W}\g{c} & \gss{W}{}{2}\g{c} & \cdots & \gss{W}{}{L_{h}-1}\g{c} \emx,
\label{eq:flatGamma}
\end{equation}
where 
\[
\g{c}=\g{B}\g{d}
\]
is the vector of \emph{pseudo-pilots} at the corresponding subcarriers. The received signal vector then becomes
\begin{eqnarray*}
\g{y} & = & \g{\Gamma}\g{h}+\g{\eta} \\ 
& = & \sum_{k=0}^{L_{h}-1}\gss{W}{}{k}\g{c}h(k) +\g{\eta} \\
& = & \left[\sum_{k=0}^{L_{h}-1}\gss{W}{}{k}h(k)\right]\g{c}+\g{\eta} \\
& = & \diag(\g{H})\g{c}+\g{\eta} \\
& = & \underbrace{\diag(\g{c})}_{\sbm{T}}\g{H}+\g{\eta}
\end{eqnarray*}
or
\begin{equation}
\g{y}=\g{T}\g{H}+\g{\eta} \label{eq:y=TH+e}
\end{equation}
with $\g{H}$ being the vector of the CFR values at the subcarrier frequencies and $\g{T}$ the diagonal matrix with the pseudo-pilots at its main diagonal (as in~\cite{kbrhk11}). 

\medskip

\noindent
\emph{Remark.}\\
In such a scenario, and for the 2-symbol preambles of Section~\ref{sec:longer}, there also holds that $\gss{G}{k}{(1,2)}=\gss{A}{}{+}$ and $\gss{G}{k}{(2,1)}=\gss{A}{}{-}$ for all $k$. Then one can readily show (using the fact that $\g{S}\gss{W}{}{k}=\gss{W}{}{k}\g{S}$) that $\g{\tilde{\bar{\Gamma}}}$ can be expressed as
\begin{equation}
\g{\tilde{\bar{\Gamma}}}=
\bmx{ccccc} \g{\bar{c}} & \g{\bar{W}}\g{\bar{c}} & \gss{\bar{W}}{}{2}\g{\bar{c}} & \cdots & \gss{\bar{W}}{}{L_{h}-1}\g{\bar{c}} \emx,
\end{equation}
with 
\[
\g{\bar{c}}=\g{\bar{B}}\g{\bar{d}}\equiv \bmx{c} \gss{\bar{c}}{1}{} \\ \gss{\bar{c}}{2}{} \emx
\]
where $\gss{\bar{c}}{q}{}$, $q=1,2$, are the pseudo-pilot vectors at the corresponding time instants, and 
\[
\g{\bar{W}}=\gss{I}{2}{}\otimes \g{W}
\]
In a manner analogous to the previous analysis, this results in the following input/output equation:
\begin{equation}
\g{\bar{y}}=\g{\bar{T}}\g{H}+\g{\bar{\eta}}, \label{eq:ybar=TH+e}
\end{equation}
where 
\[
\g{\bar{T}}=\bmx{c} \diag(\gss{\bar{c}}{1}{}) \\ \diag(\gss{\bar{c}}{2}{}) \emx=\diag(\g{\bar{c}})\bmx{c} \gss{I}{M}{} \\ \gss{I}{M}{} \emx
\]
Although this resembles~\pref{eq:y=TH+e} a lot, its analysis would be much more difficult. The reason is that $\g{\bar{T}}$ is not square anymore and the noise term, $\g{\bar{\eta}}$, is colored according to~\pref{eq:Chbar}. The rest of this appendix will focus on the single pilot symbol case. 

\subsection{Sparse Preambles}
\label{sec:flat_sparse}

The above results on optimal sparse preamble design carry over to this special case unchanged. Indeed, an optimal solution is provided by equispaced and equipowered pilot tones, as proved in~\cite{kkrt10}. The resulting minimum MSE is given by~\pref{eq:sparseMMSE} with all $\alpha_{k}$'s equal to 1, that is:
\[
\mathrm{MSE}=\frac{L_{h}\sigma^{2}}{\mathcal{E}}=\frac{1}{\mathrm{SNR}_{\mathrm{sbc}}}
\]
with $\mathrm{SNR}_{\mathrm{sbc}}=\frac{\mathcal{E}/L_{h}}{\sigma^{2}}$ denoting the SNR \emph{per subcarrier}. In the frequency domain, $\mathrm{MSE}=\frac{M}{\mathrm{SNR}_{\mathrm{sbc}}}$, which agrees with the corresponding result of~\cite{kkrt10}. 

With $\g{\alpha}=\gss{1}{L_{h}}{}$, the procedure outlined in~\pref{eq:y/a}, \pref{eq:H/a} reduces to the well known operation followed (in both CP-OFDM and FBMC/OQAM) with sparse preambles, namely estimating the CFR at the pilot tones first (dividing the received signal with the pilots) and then, through IFFT, computing an estimate for the CIR~\cite{kkrt10}. 

\subsection{Full Preambles}
\label{sec:flat_full}

The IAM methods estimate the CFR from~\pref{eq:y=TH+e} as
\[
\g{\hat{H}}=\gss{T}{}{-1}\g{y}
\]
and hence, subject to the validity of that input/output equation the corresponding MSE for CFR estimation (frequency domain MSE) is given by
\begin{eqnarray*}
\mathrm{MSE} & = & \mathrm{tr}\{\gss{T}{}{-1}\gss{C}{\sbm{\eta}}{}\gss{T}{}{-H}\} \\
& = & \sigma^{2}\sum_{m=1}^{M}\frac{1}{|c_{m}|^{2}},
\end{eqnarray*}
where $c_{m}=(\g{B}\g{d})_{m}$, 
and is to be minimized subject to the constraint~\pref{eq:aBa}. It is thus desired to maximize all $|c_{m}|^{2}$'s, hence the optimization problem can be equivalently casted as the following generalized eigenvalue problem
\[
\max_{\sbm{d}}\frac{\gss{d}{}{H}\gss{B}{}{2}\g{d}}{\gss{d}{}{H}\g{B}\g{d}},
\]
which can in turn be expressed as 
\[
\max_{\sbm{\tilde{d}}} \frac{\gss{\tilde{d}}{}{H}\g{\Lambda}\g{\tilde{d}}}{\gss{\tilde{d}}{}{H}\g{\tilde{d}}}
\]
The latter problem obviously is solved by the principal eigenvector of the matrix $\g{\Lambda}$, which in turn contains on its main diagonal the eigenvalues of
$\g{B}$. Accoding to~\cite[Lemma~1]{kbrhk11}, the maximum eigenvalue of $\g{B}$ equals $1+2\beta$ and is of multiplicity one. Hence the sought for $\g{\tilde{d}}$ vector is 
\[
\g{\tilde{d}}=\sqrt{\mathcal{E}}\gss{e}{\max}{},
\]
where $\gss{e}{\max}{}$ is the $M$-vector with all zeros except for a one at the position corresponding to the position of $1+2\beta$ at the main diagonal of $\Lambda$. The solution vector is properly normalized so that to satisfy~\pref{eq:aLa}. 
The solution for $\g{d}$ is thus given by
\begin{eqnarray*}
\g{d} & = & \sqrt{\mathcal{E}}\g{F}\gss{\Lambda}{}{-1/2}\gss{e}{\max}{} \\
& = & \sqrt{\frac{\mathcal{E}}{1+2\beta}}\gss{f}{\max}{},
\end{eqnarray*}
where $\gss{f}{\max}{}$ is the corresponding column of the $\g{F}$ matrix. It is readily verified that the latter equals 
\[
\gss{f}{\max}{}=\bmx{ccccccccccccc} 1 & -j & -1 & j & 1 & -j & -1 & j & \cdots & 1 & -j & -1 & j \emx^{T}/\sqrt{M}
\]
and hence
\[
\g{d}=\sqrt{\frac{\mathcal{E}}{M(1+2\beta)}}(\gss{1}{M/4}{}\otimes \bmx{cccc} 1 & -j & -1 & j \emx^{T})
\]
These results are in line (though derived through a different approach here) with those for optimal IAM preamble derived in~\cite{kbrhk11} and reported in~\cite{kkrt13} (see IAM-C method). 
The resulting (frequency domain) MSE can be calculated by first noting that the corresponding $\g{c}$ satisfies
\[
\g{c}\odot \gss{c}{}{*} = \frac{\mathcal{E}(1+2\beta)}{M}\gss{1}{M}{},
\]
where $\odot$ is the entry-wise product, 
and hence 
\begin{eqnarray*}
\mathrm{MSE} & = & \sigma^{2}\sum_{m=1}^{M}\frac{1}{|c_{m}|^{2}}=\frac{\sigma^{2}M^{2}}{\mathcal{E}(1+2\beta)} \\
& = & \frac{1}{\mathrm{SNR}_{\mathrm{sbc}}}\cdot\frac{M}{1+2\beta},
\end{eqnarray*}
where, again, $\mathrm{SNR}_{\mathrm{sbc}}=\frac{\mathcal{E}/M}{\sigma^{2}}$ is the per subcarrier SNR. 

\medskip

\begin{figure}
\begin{center}
\includegraphics[width=12cm]{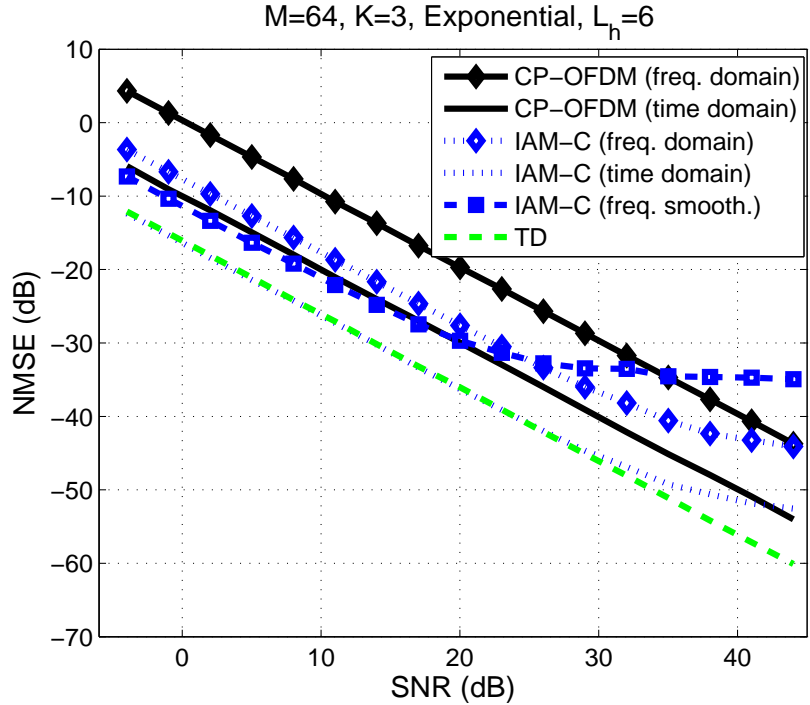}
\end{center}
  \caption{As in Fig.~\ref{fig:MSEfull}(a), including IAM-C with BLUE frequency smoothing (see eq.~\pref{eq:BLUE}).}
  \label{fig:MSEfullfs}
\end{figure}

\noindent
\emph{Remark.}
Another way to \emph{smooth} the obtained CFR values estimates besides the DFT interpolation mentioned (and tested) above is also suggested in~\cite{krbh13} (see also~\cite{kqgwj13} for a related, yet less detailed study) and relies on the assumption of the CFR being locally constant, i.e., that $H(m)$ is (almost) invariant over a certain interval around the $m$th subcarrier frequency. This is another way to say that the channel coherence bandwidth is large compared to the subcarrier spacing. The shortest such interval, namely the one covering the immediate neighbors in the subcarrier domain, will be assumed here.
The above imply that one can write (for the $m$th subcarrier)
\[
\hat{H}(p)=H(m)+\frac{\eta_{p,1}}{c_{p}}, \mbox{\ \ for\ } p\in\{m-1,m,m+1\}
\]
or 
\begin{equation}
\gss{\hat{H}}{m}{}=H(m)\gss{1}{3}{}+\gss{T}{m}{-1}\gss{\eta}{m}{}
\label{eq:window3}
\end{equation}
with obvious definitions. The \emph{best linear unbiased estimate (BLUE)} of $H(m)$ based on these ``data'' is then given by~\cite{k93}
\[
\hat{\hat{H}}(m)=\frac{\gss{1}{3}{T}\gss{C}{m}{-1}\gss{\hat{H}}{m}{}}{\gss{1}{3}{T}\gss{C}{m}{-1}\gss{1}{3}{}}
\]
with 
\[
\gss{C}{m}{}=\gss{T}{m}{-1}\gss{C}{\sbm{\eta}_{m}}{}\gss{T}{m}{-H}
\]
being the covariance of the estimatior error in~\pref{eq:window3}. Clearly,
\[
\gss{C}{m}{-1}=\frac{1}{\sigma^{2}}\gss{T}{m}{H}\gssa{B}{}{-1}{m-1:m+1,m-1:m+1}\gss{T}{m}{}
\]
where 
\[
\ga{B}{m-1:m+1,m-1:m+1}=\bmx{ccc} 1 & j\beta & 0 \\ -j\beta & 1 & j\beta \\ 0 & -j\beta & 1 \emx\equiv \gss{B}{3}{}
\]
Finally: 
\begin{equation}
\hat{\hat{H}}(m)=\frac{\gss{c}{m}{H}\gss{B}{3}{-1}(\gss{c}{m}{}\odot \gss{\hat{H}}{m}{})}{\gss{c}{H}{m}\gss{B}{3}{-1}\gss{c}{m}{}}
\label{eq:BLUE}
\end{equation}
again with obvious definitions. 
This estimator is also derived in~\cite{krbh13}, yet without the correlation of the noise components among subcarriers taken into account. The result is~\pref{eq:BLUE} with $\gss{B}{3}{-1}$ replaced by the identity matrix and is given by~\cite[eq.~(5)]{krbh13}
\begin{equation}
\hat{\hat{H}}(m)=\frac{\sum_{p=m-1}^{m+1}|c_{p}|^{2}\hat{H}(p)}{\sum_{p=m-1}^{m+1}|c_{p}|^{2}}
\label{eq:BLUEw}
\end{equation}
The filtering defined in~\pref{eq:BLUE} was applied in the example of Fig.~\ref{fig:MSEfull}(a) and the result is shown in Fig.~\ref{fig:MSEfullfs}. 
The filter of~\pref{eq:BLUEw} was also tested and gave similar results.
Observe that frequency smoothing offers a significant improvement in estimation accuracy at low to medium SNRs. However, at higher SNRs, the error floor is now more severe than before, due to the failure of the assumption of equal $H(p)$ for $p=m-1$ to $m+1$. To cope with this, a windowing that weighs more the central frequency of the frequencies involved can be applied here as well, in accordance to~\cite[eq.~(8)]{krbh13}. This modification can provide more accurate estimates, particularly if the unbiasedness constraint is relaxed~\cite{krbh13}.

\end{document}